\documentstyle[aps,amssymb,preprint,tighten,epsfig]{revtex}

\begin{document}
\draft
\title{
Influence of Long-range Disorder on Electron Motion 
in Two Dimensions}
\author{D. Taras-Semchuk$^1$ and K. B. Efetov$^{1,2}$}
\address{$^1$Theoretische Physik III, Ruhr-Universit\"{a}t\\
Bochum, Universit\"{a}tsstr. 150, 44780 Bochum, Germany\\
$^2$L. D. Landau Institute for Theoretical Physics, 117940 Moscow, Russia}
\date{\today}
\maketitle

\begin{abstract}
We consider a two-dimensional electron gas with long range
disorder. Assuming
that time reversal symmetry is broken either by an external magnetic field
or, as in the case of a delta-correlated 
random magnetic field, by the disorder itself, we derive a supermatrix $%
\sigma $-model. As an intermediate step, we provide a microscopic
derivation of the ballistic
$\sigma$-model, and find that certain corrections to its usual form
may become important.
We then integrate out degrees of freedom corresponding to
short length scales to derive a low-energy supermatrix $%
\sigma $-model. We find an extra term in the free energy 
that couples to the correlator of local currents. Use of a proper 
ultraviolet regularisation procedure that preserves gauge invariance 
indicates that the contribution of the extra term seems finally to
become irrelevant. Within the scope of our analyis, 
we therefore do not find any deviation of the scaling behaviour
of the delta-correlated random magnetic field model from that of the
conventional unitary ensemble. 
We then generalize the discussion to include models
of even longer-ranged disorder, plus short-range disorder. When the 
disorder is sufficiently long-ranged that 
the local currents become delta-correlated, a new
term appears in the free energy that does give rise to logarithmic
corrections to the conductivity. A renormalisation group analysis of
the free energy yields a scaling form for the diffusion coefficient
which contains both a positive correction, that represents classical
superdiffusion, and a negative correction, which is the usual weak
localization correction. The fact that both corrections are of the
same order and opposite sign leads to the interesting possibility of a
quantum phase transition at weak disorder in two dimensions, 
tuned by the relative strengths of the short and long range disorder.
\end{abstract}


\section{Introduction}
Since the appearance of the scaling theory of disordered metals \cite
{Wegner76,gang4,Wegner}, the theory of localization in $2D$ systems has 
become well established. Of course, this phenomenon depends on symmetries.
The addition of spin-orbital impurities leads to the
 violation of central symmetry and hence
a situation when the resistivity
vanishes at small frequencies. 
At the same time, the delocalizing effect of 
a magnetic field is apparent in a number of
well-known phenomena, a natural example being the integer quantum Hall
effect in $2D$ \cite{laughlin,halperin} (for a review see e.g. Ref.\cite
{Janssen}). Another example is provided by the random flux model which
describes, for example, electron hopping on a bipartite lattice structure
with link disorder \cite{Gade,Altland}. In this model a tendency towards
delocalization is displayed as the band center is approached due to the
existence of a chiral symmetry at the band center \cite{Gade,m+wang}.

A question of long-standing interest in this context 
has been the influence of a random,
static magnetic field on a two-dimensional electron gas. This is an example
of a disordered system with a broken time reversal invariance, where
the effects
of the magnetic field can become very complicated. Interest in this problem has
been stimulated by its relevance to a variety of experimental situations.
The solution of the problem might help to clarify the behavior of composite
fermions for the fractional quantum Hall effect near half-filling \cite
{KalHalp}. In this model, the interacting electron liquid is replaced by a
Fermi gas of quasiparticles, each carrying two flux quanta of a fictitious
magnetic field. While the Chern-Simons field exactly cancels the external
field at half-filling, variations in the electron density due to screening
of the impurity potential leads to fluctuations in the effective magnetic
field around the zero value. A similar model applies also in the gauge-field
description \cite{IofNag} of doped Mott insulators, where slow fluctuations
in the gauge field may be approximated by a static random field. Further
experimental realizations have involved the introduction of a random magnet
field (with non-zero mean) onto a high-mobility 2D semiconductor through an
overlayer containing randomly pinned flux vortices \cite{Geim} or type-I
superconducting grains \cite{Smith} or through a nearby permanent magnet 
\cite{Mancoff}.

A theoretical analysis of the random magnetic field (RMF) problem is made
technically difficult due to the spatially long-ranged nature of the vector
potential correlations, which result even for short-ranged magnetic field
correlations. For example, a straightforward application of perturbation
theory \cite{AltIof,KhvMesh} readily demonstrates the appearance of infrared
divergencies in the Born approximation for the single-particle Green
function. Consequently a wide variety of alternative techniques have been
applied to this problem. A real-space path integral representation has been
introduced by Altshuler and Ioffe \cite{AltIof} and Altshuler {\em et al}. 
\cite{EAlt}, while the eikonal \cite{Khvesh} and related paraxial \cite
{Shelankov} approximations have also been employed. If the correlation
length, $d$, of the random magnetic field is sufficiently large that $d\gg
l,1/k_{F}$, where $l$ is the single-particle mean free path and $k_{F}$ is
the Fermi momentum, then a ``classical'' regime is reached in which the
contribution of classical memory effects has been shown to be significant 
\cite{nonMark,Wilke}. If in addition the RMF is sufficiently strong that $d$ is
greater than the average cyclotron radius, then a description of electron
transport in terms of percolation between a network of `snake'-states at
zero B-field contours becomes appropriate \cite{percol,Lee,Zirnbauer}.

Numerical investigation of the RMF model (equivalent to the random
flux model away from the band center) has continued without apparent
consensus, mainly due to the great difficulty of distinguishing between
delocalization and localization of states with the very large localization
length that is typical for two dimensions at large conductances. Conclusions
for this model are divided between localization of all states \cite
{Sugiyama,Leeloc,Batsch}, the existence of a critical region \cite
{Kalmeyer,Avishai,Kawa,Liu,Yang,Sheng}, and the localization of all states
except at precisely zero energy \cite{Furusaki,m+wang}. 
It is also of interest that, even for the
case of one dimension, recent analytical \cite{Izrailev} and numerical \cite
{Moura} work supports the existence of a metal-insulator transition in the
presence of sufficiently long-ranged disorder correlations.

In this paper, we study the $2D$ electron gas with long range disorder. We
assume that the time reversal symmetry is broken. Our approach applies
both to the RMF model and also to a 
model with a long range, potential disorder and a constant
magnetic field. The latter model is very close to models with short-range
disorder with broken time-reversal invariance (unitary ensemble) for which
the localization of all states is well established. The leading order
weak localization correction to the conductivity appears, for short range
disorder, at two-loop order and is negative\cite{Wegner,Brezin,Efetov}. While
this result has been also derived from conventional diagrammatics~\cite
{hikami1}, an equivalent and more convenient procedure is provided by field
theoretical methods based on a nonlinear $\sigma $-model. In addition, the
use of the $\sigma $-models allows one to prove the existence of the
renormalization group (RG) and go beyond perturbation theory. 

Aronov {\em et al}.\cite{Aronov} have already provided a 
generalization of the standard field-theoretical approach\cite
{Wegner,Efetov} to the model of a delta-correlated RMF, in deriving
an
appropriate form of a supersymmetric $\sigma $-model. They found that the $%
\sigma $-model was identical to the one derived previously for short-range
disordered systems with broken time-reversal invariance (the unitary
ensemble)\cite{Efetov}, and therefore concluded that all states are
localized as for the unitary ensemble. In an earlier paper, Zhang and Arovas 
\cite{arovas} had proposed the existence of an additional term in the $%
\sigma $-model originating from logarithmic interaction of topological
density or, equivalently, current-carrying edge states around magnetic
domains; this term was argued to lead to a Kosterlitz-Thouless transition
from the localized to extended states with power-law correlations.
Unfortunately the calculation of Ref. \cite{arovas} relies on an
incorrect relation between a correlator of Hall conductivities and the
average longitudinal conductivity, and the additional term was not found in
the analysis of Aronov {\em et al}. \cite{Aronov}. 
As concerns the models in a constant
magnetic field with an arbitrary range of disorder correlations, a
conventional $\sigma $-model, with a classical diffusion coefficient
appropriate to the correlations, has been derived\cite{woelfle} using
similar methods to the work of Aronov {\em et al}.

Even so, the question has so far remained open 
of whether the new term written by Zhang and
Arovas\cite{arovas} 
can exist in principle and, if so, what would be the possible
consequences. We examine
this possibility 
more carefully in this paper. First we focus on the case of a
delta-correlated RMF model. For this model, 
we provide a derivation of the
appropriate $\sigma $-model, that is somewhat more complete than the 
original work of Aronov {\em et al}.\cite{Aronov}. Our derivation
demonstrates the appearance of certain technical
subtleties that were not realised in the original work of Ref.~\cite{Aronov}.  
This work is an extension of calculations presented
previously in a short letter\cite{TEfetov}. As a first step we derive 
a form of the free energy that has appeared previously under the name
of the ballistic $\sigma$-model\cite{Muz,Andreev}. In fact our
derivation (also presented in the Letter\cite{TEfetov})
represents the first formally justified derivation of the
ballistic $\sigma$-model, since the long-ranged nature of the disorder 
provides a small parameter (the small ratio of the single-particle and 
transport lifetimes) to allow a controlled separation of modes and 
expansion of the free energy, in contrast to previous derivations 
\cite{Muz,Andreev}. Indeed this approach allows us to go further and
derive further terms to the usual ballistic $\sigma$-model that become 
important for shorter-range disorder. We remark that our derivation 
of the ballistic $\sigma$-model is valid for long-range potential
disorder as well as for vector potential disorder, with minor modifications.

Integration over the degrees of freedom in this model that correspond
to length-scales shorter than the transport mean free path
 allows us to derive the final
form of the free energy that is applicable for the description of 
low-energy transport. Along with the usual terms of the unitary
$\sigma$-model, we find an extra term which couples to the
current-current correlator. This term appears also for models of
long-range potential disorder, as well as vector potential disorder, 
as long as time-reversal symmetry is completely broken. 
The new term is similar to that written by Zhang and Arovas, except
for the appearance of 
an additional factor $\Pi({\bf q})$ that represents the
current-current correlator. 

For evaluation of the quantity $\Pi({\bf q})$, it is necessary to
employ an ultraviolet regularization procedure which should, as
shown recently by Gornyi {\em et al}.\cite{Gornyi}, be formulated
carefully in order to preserve gauge invariance.
The use of this procedure then indicates that the factor $\Pi({\bf
  q})$ should vanish as ${\bf q}\to 0$, as confirmed by diagrammatic 
arguments. This implies that, within the
scope of our analysis, the new term 
remains irrelevant for the delta-correlated RMF model in the
calculation of corrections to the conductivity that are logarithmic in
frequency, although it may still lead to higher order corrections
known as memory effects (we remark that our earlier
letter\cite{TEfetov} employed a more naive regularisation procedure
which led to incorrect conclusions). 

As a result we do not find
evidence for deviation of the scaling behaviour
of the delta-correlated random magnetic field model from that of the
conventional unitary ensemble.  As such our conclusions coincide
with those of the original work of
Aronov {\em et al}., although for
considerably more subtle reasons than were originally realised in that 
work (which remains a partial analysis since they neglect higher
harmonics of the $Q$-matrix). Even so, certain questions remain, 
concerning the role of the massive modes of the field theory, which
deserve further study before a definitive conclusion can be drawn for
this problem. 

We then generalize the argument to consider the possibility of
longer-ranged disorder. 
We derive the free energy for a model of
short-range disorder plus longer-ranged disorder, and showed how the
long-range disorder couples to the free energy via a Wess-Zumino
term. If the long-range disorder is sufficiently
long-ranged that correlator of local currents remains finite in the
limit of ${\bf q}\to 0$, then a new term appears in the free energy
which is responsible for new corrections to the
conductivity that are logarithmic in frequency. For this model,
the new term does takes the
same form as the one originally written by Zhang and
Arovas\cite{arovas} (although the latter work does not contain a
correct microscopic justification). As follows from the preceding
discussion, for this term to exist the magnetic field correlations
must be longer-ranged than delta-correlated: instead, the Fourier
transform of the correlator
$\langle B({\bf r}) B ({\bf r}^{\prime})\rangle$ must diverge as $1/q^2$
as ${\bf q}\to 0$ , where $B$ is the magnetic field.

At the classical level, this free energy represents particle motion in 
a field of random velocities: similar classical models have already
been considered\cite{Derrida,KravLern,Fisher}, assuming the velocities
to be delta-correlated. Indeed, Kravtsov 
{\it et al}.\cite{KravLern} and Fisher {\em et al}.\cite{Fisher}
have demonstrated, via a renormalization group analysis, that particle
propagation in this purely classical problem (the
``diffusion-advection model'') obeys a  ``superdiffusion''.
The electron gas with the long-range disorder represents a quantum
generalisation of this model, in which quantum coherence phenomena are 
also present and interplay with the above classical effects. 

By performing a RG analysis of the free
energy, we find that two competing contributions to the scaling form
for the diffusion
coefficient appear. The positive contribution corresponds to classical
superdiffusion in the presence of long-ranged currents, while the
negative contribution is the quantum mechanical, weak localization correction
for the unitary ensemble. Since the classical and
quantum contributions to the conductivity are of opposite sign and 
potentially of the same
order, we have the interesting possibility of a quantum phase 
transition in two dimensions at weak
disorder, tuned by the relative disorder strengths. 

The plan for this paper is as follows. In section \ref{sec:Model}, we
introduce the model for the delta-correlated RMF. 
In section \ref{sec:functional}, we derive
the free-energy functional for this model in terms of a supermatrix $\sigma $%
-model. This involves deriving the ballistic $\sigma$-model
miscroscopically for this problem. 
We then simplify the free energy by
integrating out degrees of freedom associated with length-scales
shorter than the transport mean free path. The derivation is
applicable with minor changes to a model with a potential long range
disorder and a constant magnetic field. 
In section \ref{sec:pert}, we present direct perturbation theory
calculations of the conductivity and the current-current correlator.
In section \ref{sec:superdiff}, we review classical models of
diffusion in a field of random velocities, and 
consider an electron gas in the presence of both short-range and 
longer-ranged disorder. We show how the local current couples to the
free energy through a Wess-Zumino term, and how averaging over this
term leads to a term similar to that written by Zhang and
Arovas\cite{arovas}. In section \ref{sec:RG} we subject the free
energy to a renormalisation group analysis and derive the scaling form 
for the diffusion coefficient. This demonstrates the appearance of a
positive contribution to the scaling form for the
conductivity in competition with the usual
negative, weak localization contribution.
Section \ref{sec:discuss} concludes with a summary and discussion. Some 
technical details are presented in the Appendix.

\section{Random Magnetic Field Model}

\label{sec:Model}

As we have mentioned in the Introduction, the considerations below 
are applicable
both to a RMF model, and to a model with long range potential impurities
and a constant magnetic field (to break time-reversal symmetry). 
While the derivation of the $\sigma $-model is
similar for the two types of model, to be specific we carry out 
calculations for
the RMF model. We write the Hamiltonian as follows, 
\begin{equation}
{\cal H}({\bf r})=\frac{1}{2m}\left( -i\nabla _{{\bf r}}-\frac{e}{c}{\bf A}(%
{\bf r})\right) ^{2}-\epsilon _{F}  \label{a1}
\end{equation}
where $e$ and $m$ are the electron charge and mass, $c$ is the 
velocity of light, $\epsilon _{F}$ is the Fermi energy and ${\bf A}$ is the vector
potential. We focus below on the case
where the magnetic field, ${\bf B}=\nabla \times {\bf A}$, is
 delta-correlated, 
\begin{equation}
\langle B({\bf r})B({\bf r}^{\prime })\rangle =2\left( \frac{mcv_{F}}{e}%
\right) ^{2}\gamma \,\delta ({\bf r}-{\bf r}^{\prime }),  \label{a1b}
\end{equation}
with a strength characterized by the dimensionless parameter $\gamma $
(and with $%
\langle B\rangle =0$), although one may as easily 
consider a more general case with a
finite range of the correlations. We choose the London gauge for the vector
potential ${\bf A}$ 
\begin{equation}
div{\bf A=}0,\text{ \ }{\bf A}_{n}=0  \label{a1a}
\end{equation}
where ${\bf A}_{n}$ is the component taken at, and perpendicular to,
the surface of the sample.
The sample may be either truly two-dimensional, with the thickness of
a single atomic layer, or 
simply have a two-dimensional geometry. While we write
explicit formulae for the former case, extension to thicker samples is
trivial.

The correlations of the vector potential ${\bf A}$ corresponding Eqs.~(\ref
{a1b}) and (\ref{a1a}) are long-ranged and we choose them in the form 
\begin{eqnarray}
\langle A^{i}({\bf r})A^{j}({\bf r}^{\prime })\rangle  &=&\frac{2m^{2}c^{2}}{%
e^{2}}V^{ij}({\bf r}-{\bf r}^{\prime }),  \nonumber \\
\quad V^{ij}({\bf q}) &=&v_{F}^{2}\gamma \frac{1}{(q^{2}+\kappa
^{2}p_{F}^{2})}\left( \delta _{ij}-\frac{q^{i}q^{j}}{q^{2}}\right) .
\label{a2}
\end{eqnarray}
The correlator $V^{ij}$ shown here is transversal, that is, 
\[
\sum_{i}\frac{\partial }{\partial r_{i}}V^{ij}=0,
\]
which corresponds to the transversality of the vector potential ${\bf A}$,
Eq. (\ref{a1a}). The parameter $\kappa \ll 1$ is a cutoff that renders
finite the otherwise-infinite range of the disorder correlations. The $%
\delta $-correlated fluctuations of the magnetic field correspond to the
limit $\kappa \rightarrow 0$. The cutoff $\kappa $ appears \cite
{KhvMesh,Aronov} in perturbation theory for the infrared-divergent
single-particle lifetime. For example, the simple Born approximation yields
for the mean free time 
\begin{eqnarray}
\tau _{{\rm BA}}^{-1} &=&4\epsilon _{F}\gamma \int_{0}^{2\pi }\frac{d\varphi 
}{2\pi }\frac{\cos ^{2}(\varphi /2)}{4\sin ^{2}(\varphi /2)+\kappa ^{2}} 
\nonumber \\
&\simeq &2\epsilon _{F}\gamma /\kappa .  \label{tauBA}
\end{eqnarray}
The self-consistent Born approximation (SCBA) displays a weaker divergency, 
\begin{mathletters}
\begin{eqnarray}
\tau _{{\rm SCBA}}^{-1} &=&\frac{8}{\pi }\epsilon _{F}\gamma \tau _{{\rm SCBA%
}}\int_{0}^{2\pi }\frac{d\varphi }{2\pi }\int_{-\infty }^{\infty }d\xi \sin
^{2}\varphi \times   \nonumber \\
&&\hspace{-1.2cm}\frac{1}{(4\sin ^{2}(\varphi /2)+\xi ^{2}/\epsilon
_{F}^{2})(4\sin ^{2}(\varphi /2)+\xi ^{2}/\epsilon _{F}^{2}+\kappa ^{2})},
\label{tauSCBAa} \\
\tau _{{\rm SCBA}}^{-1} &\simeq &2\epsilon _{F}(\gamma \ln (1/\kappa )/\pi
)^{1/2},  \label{tauSCBA}
\end{eqnarray}
due to off-shell contributions in the collision integral, Eq.~(\ref{tauSCBAa}%
). The inverse transport time $\tau _{{\rm tr}}^{-1}$ remains convergent
however in the limit of $\kappa \rightarrow 0$, 
\end{mathletters}
\begin{eqnarray}
\tau _{{\rm tr}}^{-1}= &&\epsilon _{F}\gamma \int_{0}^{2\pi }\frac{d\varphi 
}{2\pi }(1-\cos (\varphi ))\cot ^{2}(\varphi /2)  \nonumber \\
&=&\epsilon _{F}\gamma .  \label{tautr}
\end{eqnarray}
Although we focus on the above choice of correlator $V^{ij}({\bf q})$ in
this paper, we emphasize again that our method remains valid for an
arbitrary form of disorder correlations, with minor modifications for scalar
rather than vector potential disorder.

\section{Free Energy Functional}

\label{sec:functional} In this section we derive the free energy functional
for the 2D electron gas in the RMF. We employ the supersymmetry method which
by now has been extensively developed \cite{Efetov} as a method for the
exact evaluation of spectral and wave function properties of metals with
delta-correlated disorder. In the conventional case the free energy
functional takes the form of a nonlinear $\sigma $-model containing an $%
8\times 8$, position-dependent $Q$-matrix. In order to generalize the method
to allow for long-ranged disorder correlations, it becomes necessary to
employ a $Q$-matrix that depends on two position variables, rather than one.
Following a Fourier transformation and a harmonic expansion around the Fermi
surface, the $Q$-matrix may be represented as depending on a position
variable and an angular harmonic index.

The degrees of freedom contained in the non-zero harmonics of the $Q$-matrix
are weakly massive for a long range disorder and we integrate them out
rather than neglect them. 
Although the coupling of the non-zero harmonics to
the zeroth harmonic is weak, integration over the nonzero harmonics
requires some care to determine the possible 
renormalization of the effective functional for the zeroth harmonic. 
One
consequence of this renormalization, due to the first harmonics only, amounts
to the inclusion of two-particle vertex corrections to the single-particle
lifetime, $\tau $, leading to its replacement in the free energy by the
transport relaxation time, $\tau _{{\rm tr}}$. This was the main conclusion
of Ref. \cite{woelfle}, where arbitrary range potential impurities were
considered, and of Ref. \cite{Aronov}, devoted to study of the RMF problem.

We 
find however a further difficulty associated with the integration over non-zero
harmonics that was not examined in the previous
works\cite{woelfle,Aronov}. 
Namely, additional terms
arise in the free energy whose evaluation requires a careful integration
over {\em higher} non-zero harmonics, rather than only the first harmonic as
considered in the approach of Refs.~\cite{woelfle,Aronov}. While the
contribution of these 
terms seem finally to be irrelevant to computation of logarithmic
corrections to the conductivity, as we discuss below, the
proper demonstration of this fact is far from trivial and beyond the
original analysis of Ref.\cite{Aronov}. 

In order to properly include the potential 
contribution of the higher harmonics, we derive as an intermediate step a
free energy that is similar to the ``ballistic $\sigma $-model'' of
Muzykantskii and Khmelnitskii \cite{Muz}. The latter model itself represents
a generalization of the diffusive $\sigma $-model to the case of ballistic
disorder, where the typical energy scale of $Q$-fluctuations may be as large
as the inverse scattering time. In fact we also derive certain additional
terms to their form of the free energy that become relevant for
shorter-range disorder. 

Employing standard methods\cite{Efetov}, we introduce a supersymmetric $\psi 
$-field that contains eight components, corresponding to fermion/boson,
advanced/retarded and time-reversed sectors. An averaged product of Green's
functions may then be expressed in terms of a functional integral weighted
by a Lagrangian that is quadratic in the $\psi $-field: 
\begin{eqnarray}
&&\hspace{-0.5cm}\langle {\cal G}_{\epsilon -\omega /2}^{A}({\bf r},0){\cal G%
}_{\epsilon +\omega /2}^{R}(0,{\bf r})\rangle  \nonumber \\
&=&-4\int \psi _{\alpha }^{1}({\bf r})\bar{\psi}_{\alpha }^{1}(0)\psi
_{\beta }^{2}(0)\bar{\psi}_{\beta }^{2}({\bf r})e^{-{\cal L}}{\cal D}\psi 
{\cal D}\bar{\psi},  \nonumber \\
{\cal L} &=&-i\int \bar{\psi}({\bf r})\left( {\cal H}_{0}+\frac{ie}{mc}{\bf A%
}\nabla \tau _{3}+\frac{e^{2}}{2mc^{2}}A^{2}\right) \psi ({\bf r})d%
{\bf r},  \label{Lag0}
\end{eqnarray}
where 
\[
{\cal H}_{0}\equiv -\frac{\nabla ^{2}}{2m}-\epsilon -\epsilon _{F}+\frac{%
\left( \omega +i\delta \right) \Lambda }{2}, 
\]
and $\tau _{3}$ is the Pauli matrix in time-reversal space.

We now average over the vector potential, ${\bf A}$. In doing so we neglect
the term in $A^{2}$ in Eq.~(\ref{Lag0}). 
This approximation is standard and has been
used in previous analytic works on the RMF 
model\cite{Aronov,miller}. Furthermore, 
for the case of long range potential disorder, for which our general
method is valid, $A^2$-like
terms are absent anyway. As for the question of gauge invariance, note 
that our derivation has assumed
already a choice of gauge through the form of the vector potential
correlations (see section \ref{sec:Model}). 
Nevertheless Appendix \ref{app:asquare}
contains a discussion of how the $A^2$ term may be handled more
carefully and found to be negligible.

The term linear in ${\bf A}$ induces after the averaging a term
that is quartic in the $\psi $-fields and, in contrast to the case of
short-range impurities, non-local in position: 
\[
{\cal L}_{{\rm int}}=\int \left( \bar{\psi}({\bf r})\tau _{3}\nabla _{{\bf r}%
}^{i}\psi ({\bf r}))V^{ij}({\bf r}-{\bf r}^{\prime }\right) \left( \bar{\psi}%
({\bf r}^{\prime })\tau _{3}\nabla _{{\bf r}^{\prime }}^{j}\psi ({\bf r}%
^{\prime })\right) d^{2}{\bf r}d^{2}{\bf r}^{\prime }. 
\]
Following an integration by parts, ${\cal L}$ may be rewritten 
\begin{eqnarray}
{\cal L}_{{\rm int}} &=&\int \bar{\psi}^{\alpha }({\bf r})\psi ^{\beta }(%
{\bf r}^{\prime })\tau _{3\,\beta \beta }k_{\beta \beta }\widehat{V}_{{\bf r}%
,{\bf r}^{\prime }}\bar{\psi}^{\beta }({\bf r}^{\prime })\psi ^{\alpha }(%
{\bf r})\tau _{3\,\alpha \alpha },  \nonumber \\
\widehat{V}_{{\bf r},{\bf r}^{\prime }} &\equiv &-\frac{1}{2}\sum_{ij}V^{ij}(%
{\bf r}-{\bf r}^{\prime })\left( \nabla _{{\bf r}}^{i}-\nabla _{{\bf r}%
^{\prime }}^{i}\right) \left( \nabla _{{\bf r}}^{j}-\nabla _{{\bf r}^{\prime
}}^{j}\right)  \label{Lint}
\end{eqnarray}
where $k={\rm diag}(1,-1)$ in boson-fermion space. We now decouple ${\cal L}%
_{{\rm int}}$ via an $8\times 8$ matrix $Q({\bf r},{\bf r}^{\prime })$.
While in the calculation for time reversal invariant systems\cite{Efetov},
this step requires a careful identification of the slowly varying modes, in
the case of the system with the broken time reversal invariance, all slow
modes are easily identified as corresponding to\ pairs $\psi ^{\alpha
}\left( {\bf r}\right) \bar{\psi}^{\beta }\left( {\bf r}^{\prime }\right) $.
Following an integration over the $\psi $-fields, we find the Lagrangian 
\begin{eqnarray}
{\cal L} &=&\int \left[ -\frac{1}{2}{\rm Str}\ln \left( i{\cal H}_{0}({\bf r}%
)\delta ({\bf r}-{\bf r}^{\prime })+\widehat{V}_{{\bf r},{\bf r}^{\prime }}%
\widetilde{Q}({\bf r},{\bf r}^{\prime })\right) \right.  \nonumber \\
&&\left. \mbox{}\qquad +\frac{1}{4}{\rm Str}\left( Q({\bf r},{\bf r}^{\prime
})\widehat{V}_{{\bf r},{\bf r}^{\prime }}Q({\bf r}^{\prime },{\bf r})\right) %
\right] d{\bf r}d{\bf r}^{\prime },  \label{Lag1}
\end{eqnarray}
Here $\widetilde{Q}=Q_{\shortparallel }+iQ_{\perp }$, whereas $%
Q=Q_{\shortparallel }+Q_{\perp }$ and $Q_{\shortparallel }$ ($Q_{\perp }$)
commutes (anticommutes) with $\tau _{3}$. The $Q$-matrix satisfies the
standard symmetries $Q({\bf r},{\bf r}^{\prime })=\bar{Q}({\bf r}^{\prime },%
{\bf r})=KQ^{\dagger }({\bf r}^{\prime },{\bf r})K$, where $\bar{Q}\equiv
CQ^{T}C^{T}$ and $C$ and $K$ defined as in Ref.~\cite{Efetov}.

The free energy, Eq. (\ref{Lag1}), in principle provides an exact
description of the system but it is not useful in this form as it allows for
all possible energy scales for $Q$-fluctuations. As a first step in its
simplification we search as usual for the saddle-point value of $Q$ and
perform a gradient expansion around this value to identify the free energy
for low energy fluctuations.

\subsection{Saddle-point}

To search for the saddle-point to Eq.~(\ref{Lag1}), we take $Q_{\perp }=0$
and $Q$ to depend on ${\bf r}-{\bf r}^{\prime }$ only. This means that the
cooperon ``degrees of freedom'' are suppressed and that the supermatrix $Q$ at
the saddle point is homogeneous in space. After Fourier transforming with
respect to ${\bf r}-{\bf r}^{\prime }$, the saddle-point equation reads $Q_{%
{\bf p}}=g_{{\bf p}}$, where 
\begin{equation}
\left( \epsilon -\xi _{{\bf p}}-\frac{\omega }{2}\Lambda -2i\int \frac{d^{2}%
{\bf p}_{1}}{(2\pi )^{2}}V_{{\bf p}-{\bf p}_{1}}^{ij}p_{1}^{i}p_{1}^{j}Q_{%
{\bf p}_{1}}\right) g_{{\bf p}}=-i,  \label{saddle}
\end{equation}
and $\xi _{{\bf p}}={\bf p}^{2}/(2m)-\epsilon _{F}$.

The saddle point of the Lagrangian, Eq. (\ref{Lag1}), is continuously
degenerate and the solution of Eq. (\ref{saddle}) can be written generally
in the form 
\begin{equation}
Q_{{\bf p}}=V_{{\bf p}}\Lambda _{{\bf p}}\bar{V}_{{\bf p}}\text{, \ \ }V_{%
{\bf p}}\bar{V}_{{\bf p}}=1  \label{b1}
\end{equation}
where $\Lambda _{{\bf p}}$ is a diagonal matrix depending on ${\bf p}$ in a
complicated way. The analysis simplifies however in the limits of $\kappa
\gg (\epsilon _{F}\tau )^{-1}$ (short-range disorder) and $\kappa \ll
(\epsilon _{F}\tau )^{-1}$ (long-range disorder), the latter case being
relevant here, where $\tau $ is the mean free time to be determined from the
solution. For either case we may employ the ansatz 
\begin{equation}
\Lambda _{{\bf p}}=\frac{i}{\xi _{{\bf p}}-\epsilon +i\Lambda /(2\tau )},
\label{ansatz}
\end{equation}
where $\Lambda ={\rm diag}\left( 1,1,1,1,-1,-1,-1,-1\right) $, which leads
to the condition 
\begin{equation}
\frac{\Lambda }{2\tau }=2\int \sum_{ij}V_{{\bf p}-{\bf p}%
_{1}}^{ij}p_{1}^{i}p_{1}^{j}\Lambda _{{\bf p}_{1}}\frac{d^{2}{\bf p}_{1}}{%
(2\pi )^{2}}.  \label{b2}
\end{equation}

For short-range disorder, $\kappa \gg (\epsilon _{F}\tau )^{-1}$, the
integral in Eq. (\ref{b2}) is as usual dominated by the on-shell value of $%
{\bf p}-{\bf p}_{1}$ and $\tau $ takes its simple BA value, as given by Eq.~(%
\ref{tauBA}), which is equivalent to the SCBA value in this limit.

In the opposite limit of long-ranged disorder, $\kappa \ll (\epsilon
_{F}\tau )^{-1}$, which is relevant to the RMF model, the off-shell
dependence of $V_{{\bf p}-{\bf p}^{\prime }}^{ij}$ contributes significantly
to the integral. In this limit, substituting Eq. (\ref{ansatz}) into Eq. (%
\ref{b2}) we come to Eq.~(\ref{tauSCBAa}), thus obtaining the SCBA value of $%
\tau $, given by Eq. (\ref{tauSCBA}). This value of $\tau $ is less
divergent than the simple BA value, Eq. (\ref{tauBA}). In the following we
will denote this SCBA value as simply $\tau $. At the same time we ensure
that we stay within the weak disorder limit, $\epsilon _{F}\tau \gg 1$
(equivalently, $\gamma \ll 1$). This condition restricts the space of
relevant fluctuations in the $Q$-matrix, as we see in the next section.

\subsection{Fluctuations}

\label{sec:flucts}

To proceed we expand the free energy in fluctuations of $Q$ about its
saddle-point value. In the case of short range impurities, fluctuations of
the eigenvalues of the supermatrix $Q$ are massive with the characteristic
energy $1/\tau $ and may be neglected. In this case, what 
remains is to consider massless 
(in the limit $\omega \rightarrow 0$) modes \cite{Efetov}
with energies smaller than $1/\tau $, which leads to the conventional
supermatrix $\sigma $-model.

In the models with long range disorder considered now, however, 
an additional
energy scale appears due the difference between the mean free time $\tau $
and $\tau _{{\rm tr}}$. In this case 
fluctuations separate into three types: \newline
(a) hard massive, with a mass of $\tau ^{-1}$, \newline
(b) soft massive, with a mass of $\tau _{{\rm tr}}^{-1}$, and \newline
(c) `massless', that is, with a mass vanishing in the limit $\omega
\rightarrow 0$. \newline
Type (a) fluctuations are associated with the fluctuations of the
eigenvalues of $Q$ and the Cooperons. The weak disorder condition $\epsilon
_{F}\tau \gg 1$ ensures that these fluctuations are irrelevant and hence we
may restrict attention to $Q$-matrices of the form $Q^{2}%
=1$ and $[Q,\tau _{3}]=0$. Alternatively, one can use the form of
the supermatrices $Q$ given by Eq. (\ref{b1}) with $\left[ V,\tau _{3}\right]
=0$. The effective functional for the remaining (b) and (c) modes will then
contain energy scales only much smaller than $\tau ^{-1}$. 

The above separation of scales may also be expressed in
terms of length scales: at distances exceeding $l_{{\rm tr}%
}=v_{F}\tau _{{\rm tr}}$, density relaxation in the classical limit
is described by a diffusion equation, whereas
for distances between $l=v_{F}\tau $ and $l_{{\rm tr}}$ one should use a
Boltzmann equation. The possibility to generalize the above classical
descriptions to include quantum fluctuation phenomena is afforded by 
the $\sigma $-model technique.

To derive a free energy functional describing both soft massive and massless
modes we proceed as follows. First we introduce the following Fourier
transformation for $Q({\bf r},{\bf r}^{\prime })$ with respect to ${\bf r}-%
{\bf r}^{\prime }$: 
\begin{eqnarray}
Q({\bf r},{\bf r}^{\prime })=\int \frac{d^{2}{\bf p}}{(2\pi )^{2}}Q_{{\bf p}%
}({\bf R})\exp (i\tau _{3}{\bf p}({\bf r}-{\bf r}^{\prime })), 
\label{Ftrans}
\end{eqnarray}
where ${\bf R}=({\bf r}+{\bf r}^{\prime })/2$. The appearance of the $\tau
_{3}$ factor in Eq.~(\ref{Ftrans}) ensures that $Q_{{\bf p}}({\bf R})$
satisfies the symmetries 
\[
\bar{Q}_{{\bf p}}({\bf R})=Q_{{\bf p}}({\bf R})=KQ_{{\bf p}}^{\dagger }({\bf %
R})K. 
\]
Next we introduce the following parametrization for $Q_{{\bf p}}({\bf R})$: 
\begin{eqnarray}
Q_{{\bf p}}({\bf R}) &=&U\left( {\bf R}\right) Q_{{\bf p}}^{\left( 0\right) }%
\bar{U}\left( {\bf R}\right) ,  \nonumber \\
Q_{{\bf p}}^{\left( 0\right) } &=&V_{{\bf n}}({\bf R})\Lambda _{{\bf p}}\bar{%
V}_{{\bf n}}\left( {\bf R}\right) ,  \label{param}
\end{eqnarray}
where ${\bf n}={\bf p}/|p|$. In Eq. (\ref{param}), $U\left( {\bf R}\right) $
is independent of ${\bf n}$, and $Q_{{\bf p}}^{\left( 0\right) }\left( {\bf R%
}\right) $ contains only non-zero harmonics in ${\bf n}$ around the Fermi
surface. The matrices $U$ and $V_{{\bf n}}$ obey the symmetries 
\begin{equation}
\begin{array}{rclrcl}
\bar{U}U & = & 1, & \bar{U} & = & KU^{\dagger }K, \\ 
\bar{V}_{{\bf n}}V_{{\bf n}} & = & 1, & \bar{V}_{{\bf n}} & = & KV_{{\bf n}%
}^{\dagger }K.
\end{array}
\label{Usymm}
\end{equation}

In what follows we will make expansions in deviations of the supermatrix $Q_{%
{\bf p}}^{\left( 0\right) }\left( {\bf R}\right) $ from $\Lambda _{{\bf p}}$%
. For this purpose the supermatrix $Q_{{\bf p}}^{\left( 0\right) }\left( 
{\bf R}\right) $ can be conveniently parametrized, for example, as 
\begin{equation}
Q_{{\bf p}}^{\left( 0\right) }\left( {\bf R}\right) =\Lambda _{{\bf p}%
}\left( \frac{1+iP_{{\bf n}}({\bf R})}{1-iP_{{\bf n}}({\bf R})}\right)
\label{a6}
\end{equation}
with $\int d{\bf n}P_{{\bf n}}=0$, $\overline{P}_{{\bf n}}=-P_{{\bf n}}$, $%
\left\{ P_{{\bf n}},\Lambda \right\} =0$. Both $U$ and $V_{{\bf n}}$ vary
slowly with ${\bf R}$, that is, on length scales longer than $l
=v_{F}\tau$. While $U({\bf r})$ represents the massless (c)
modes, the weakly massive modes (b) are contained in $V_{{\bf n}}$. 
It is the dependence of the supermatrix $V_{{\bf n}}$ on the 
vector ${\bf n}$ that leads to the  
gap in the spectrum of excitations of the (b) modes.

An important feature of our parametrization for $Q_{{\bf p}}\left( {\bf R}%
\right) $ in the form 
\[
Q_{{\bf p}}\left( {\bf R}\right) =T_{{\bf n}}\left( {\bf R}\right) \Lambda _{%
{\bf p}}\bar{T}_{{\bf n}}\left( {\bf R}\right) 
\]
is the separation of the rotation matrix $T_{{\bf n}}=UV_{{\bf n}}$ into the 
$U$ and $V_{{\bf n}}$ factors. This step is not only convenient for
computation but, more importantly, ensures the preservation of an original
symmetry of the initial Lagrangian, Eq.~(\ref{Lag0}). Specifically, it
ensures that the final free energy functional is invariant under global
rotations $U\left( {\bf R}\right) \rightarrow U_{0}U\left( {\bf R}\right) $
with $\bar{U}_{0}=U_{0}^{-1}$ independent of coordinates. This invariance
follows from the original invariance of the Lagrangian, Eq.~(\ref{Lag0}),
under global rotations $\psi ({\bf r})\rightarrow U_{0}\psi ({\bf r})$.

Substitution of the parametrization (\ref{param}) into expression (\ref{Lag1}%
) yields a free energy in terms of only the weakly massive (b) modes and the
massless (c) modes. Accordingly this free energy will describe fluctuations
over energy scales only much less than $\tau ^{-1}$. In order to investigate
the truly low-frequency behavior of transport coefficients, however, we need
to reduce this free energy even further by integrating over the weakly
massive (b) modes. This procedure will reduce the free energy to a form in
terms of only $U({\bf r})$, representing the massless (c) modes. In this way
we obtain the final form of the free energy that is applicable on energy
scales much less than $\tau _{{\rm tr}}^{-1}$ and so appropriate for the
description of low-energy transport. It is necessary to integrate carefully
over the (b) modes, rather than simply neglecting them, because, even though
their coupling to the (c) modes is weak, it is sufficiently complicated to
lead on integration to a potentially
non-trivial renormalization of the bare free energy
for the (c) modes. Due to the soft mass of the (b) modes it is sufficient to
apply a Gaussian approximation in $P_{{\bf n}}$; higher order terms give a
small contribution provided the inequality $\epsilon _{F}\tau _{{\rm tr}}\gg
1$ (equivalently, $\gamma \ll 1$) is fulfilled.

In principle, if the correlations of the disorder decay sufficiently 
slowly (for the
form of the correlations given by Eq.~(\ref{a2}), this applies in the
limit of $%
\kappa \rightarrow 0$), one needs to employ a cutoff at large momentum $k$ (short
distance). The need for an
ultra-violet regularization is well-known \cite{Andreev,Aleiner,Izyumov} in
the context of the ballistic $\sigma $-model and originates from the fact
that the gradient (Liouvillean) operator in the logarithm of the initial
Lagrangian, Eq.~(\ref{Lag1}), is singular in two dimensions. It turns
out that in this model 
one needs to be careful about the precise form of the
short-distance cutoff, a point 
which we discuss further below.

It is helpful at this point to make contact with the conventional
calculation for delta-correlated impurities \cite{Efetov}. To do so we
simply set $V_{{\bf n}}=1$ in Eq.~(\ref{param}) so that $Q$ becomes a
function of ${\bf R}$ only. A straightforward gradient expansion of the
Lagrangian (\ref{Lag1}) to second-order in the gradients then recovers the
conventional $\sigma $-model. It is useful also to note that,
prior to the gradient expansion, the 
Lagrangian (\ref{Lag1}) is invariant under a local
``gauge'' transformation 
\begin{equation}
U({\bf R})\rightarrow U({\bf R})h({\bf R}),\text{ \ \ \ }[h({\bf R}),\Lambda
]=0  \label{b5a}
\end{equation}
(note the term ``gauge'' is used here in a separate sense from that of the
original vector potential). As we will see later, this invariance
remains conserved 
also for the case of long range disorder and is related to conservation of
local currents.

\subsection{Gradient expansion}

\label{sec:gradient}

We proceed with the derivation of the free energy functional by
expansion of the
logarithm in the Lagrangian (\ref{Lag1}) in low energy terms. Substituting
the parametrization, Eq. (\ref{param}), into the Lagrangian (\ref{Lag1}),
and cycling factors under the supertrace, we find 
\begin{eqnarray}
{\cal L} &=&\int \left[ -\frac{1}{2}{\rm Str}\ln \left( -i\xi _{p}+\frac{%
\Lambda }{2\tau }+{\cal A}_{{\rm kin}}+{\cal A}_{{\rm coll}}+{\cal A}%
_{\omega }\right) \right.  \nonumber \\
&&\left. +\frac{1}{2}\int {\rm Str}\left( Q_{{\bf p}}^{(0)}({\bf R})\,V_{%
{\bf p}-{\bf p}_{1}}^{ij}p_{1}^{i}p_{1}^{j}\,Q_{{\bf p}_{1}}^{(0)}({\bf R}%
)\right) \frac{d^{2}{\bf p}_{1}}{(2\pi )^{2}}\right] d{\bf R}\frac{d^{2}{\bf %
p}}{(2\pi )^{2}},  \label{Lag2}
\end{eqnarray}
where 
\begin{eqnarray*}
{\cal A}_{{\rm kin}} &=&v_{F}\bar{T}_{{\bf n}}({\bf R}){\bf n}\nabla _{{\bf R%
}}T_{{\bf n}}({\bf R})\tau _{3}, \\
{\cal A}_{{\rm coll}} &=&2\int V_{{\bf p}-{\bf p_{1}}%
}^{ij}p_{1}^{i}p_{1}^{j}\left( \bar{T}_{{\bf p}}({\bf R})Q_{{\bf p}_{1}}(%
{\bf R})T_{{\bf p}}({\bf R})-\Lambda _{p_{1}}\right) \frac{d^{2}{\bf p}_{1}}{%
(2\pi )^{2}}, \\
{\cal A}_{\omega } &=&-i\left( \omega +i\delta \right) \bar{T}_{{\bf n}}(%
{\bf R})\Lambda T_{{\bf n}}({\bf R}),
\end{eqnarray*}
and $T_{{\bf n}}=UV_{{\bf n}}$. The term ${\cal A}_{{\rm kin}}$ describes
the kinetic energy, ${\cal A}_{{\rm coll}}$ is the collision integral and $%
{\cal A}_{\omega }$ is the usual frequency 
term entering the $\sigma $%
-model. Then we perform an expansion of the logarithm in the terms ${\cal A}%
_{{\rm kin}}$, ${\cal A}_{{\rm coll}}$ and ${\cal A}_{\omega }$. If we keep
to first order in all of these terms, 
we find that the contribution of ${\cal A}_{%
{\rm coll}}$ is $(-2)$ times that of the last term in Eq.~(\ref{Lag2}): in
this way we recover the usual form\cite{Muz,Andreev} of the $\sigma $-model
for ballistic disorder: 
\begin{eqnarray}
F &=&\frac{\pi \nu }{4}{\rm Str}\int \left[ \frac{1}{2}\int w({\bf n},{\bf n}%
^{\prime })(Q_{{\bf n}}({\bf r})-Q_{{\bf n}^{\prime }}({\bf r}))^{2}\frac{d%
{\bf n}d{\bf n}^{\prime }}{(2\pi )^{2}}\right.  \nonumber \\
&&\left. -2v_{F}\int \Lambda \tau _{3}\bar{T}_{{\bf n}}({\bf r}){\bf n\nabla 
}_{{\bf r}}T_{{\bf n}}({\bf r})\frac{d{\bf n}}{2\pi }+i\omega \int \Lambda
Q_{{\bf n}}({\bf r})\frac{d{\bf n}}{2\pi }\right] d{\bf r},  \label{M+K}
\end{eqnarray}
where $\nu $ is the density of states and $w({\bf n}_{1},{\bf n}_{2})$ is
defined as 
\begin{equation}
w({\bf n}_{1},{\bf n}_{2})=2\pi \nu p_{F}^{2}\sum_{ij}V_{{\bf n}_{1}-{\bf n}%
_{2}}^{ij}n_{1}^{i}n_{1}^{j}  \label{b7}
\end{equation}

so that 
\begin{eqnarray}
\frac{1}{2\tau _{{\rm BA}}} &=&\int w({\bf n},{\bf n}^{\prime })\frac{d{\bf n%
}d{\bf n}^{\prime }}{(2\pi )^{2}},  \nonumber \\
\frac{1}{2\tau _{{\rm tr}}} &=&\int w({\bf n},{\bf n}^{\prime })(1-{\bf nn}%
^{\prime })\frac{d{\bf n}d{\bf n}^{\prime }}{(2\pi )^{2}}.  \label{tauw}
\end{eqnarray}
In the limit $\kappa \rightarrow 0$ the function $w\left( {\bf n}_{1},{\bf n}%
_{2}\right) $ takes a simpler form 
\begin{equation}
w\left( {\bf n}_{1},{\bf n}_{2}\right) =\frac{v_{F}^{2}\nu \pi \gamma }{2}%
\frac{(1+{\bf n}_{1}{\bf n}_{2})}{(1-{\bf n}_{1}{\bf n}_{2})}\qquad (\kappa
\rightarrow 0)  \label{b8}
\end{equation}

Notice that the free energy $F$, Eq. (\ref{M+K}), does not contain any
divergencies in the limit of $\kappa \rightarrow 0$, since the singularity
in $w({\bf n},{\bf n}^{\prime })$ as ${\bf n}\rightarrow {\bf n}^{\prime }$,
Eq. (\ref{b8}), is compensated by the $(Q_{{\bf n}}-Q_{{\bf n}^{\prime
}})^{2}$ factor in the collision integral of Eq.~(\ref{M+K}). This is a
reflection of the fact that the free energy contains energy scales only less
than $\tau ^{-1}$ since we have eliminated the hard massive (a) modes.

We emphasize that, while the form of the 
ballistic $\sigma$-model in Eq.~(\ref{M+K}) 
has been presented previously\cite{Muz,Andreev}, the derivation here for
the case of long-ranged disorder represents the first formally
justified derivation of this model. This has been made possible by the 
separation of the single-particle and transport lifetimes, $\tau$ and
$\tau_{\rm tr}$: this leads 
to the small parameter $\tau/\tau_{\rm
  tr}\ll 1$ which separates the (a) type fluctuations from the (b) and 
(c) type fluctuations, and allows a gradient
expansion of the free energy in a controlled manner. 
The derivation remains valid
also for long-range potential disorder, instead of vector potential.
In this case, Cooperons become operative unless time-reversal symmetry
is completely broken by a constant magnetic field.

An interesting point is that, for the RMF model, the ratio 
$\tau/\tau_{\rm tr}$ formally vanishes as the cutoff $\kappa$ is taken 
to zero, at least when the SCBA value of the single-particle lifetime
$\tau$ is taken. This suggests superficially that the above derivation of 
the ballistic $\sigma$-model of Eq.~(\ref{M+K}) 
becomes exact for the RMF model
in the limit of $\kappa\to 0$, and in principle applicable on
length-scales all
the way down to the Fermi wavelength. A peculiarity of the RMF problem 
however is that it is not straightfoward to specify the precise value
of the single-particle lifetime: for example, a gauge-invariant
formulation\cite{EAlt} 
produces a value of the single-particle lifetime that is
convergent, rather than vanishing, as $\kappa\to 0$, 
at least when the average magnetic field
is nonzero. The precise value of the single-particle lifetime in the
RMF model that
couples to the (a)-mode fluctuations, and hence the value of the
shortest length-scales down to which the ballistic $\sigma$-model is
applicable for this model, remains an open question. 

A further strength of the derivation of the ballistic $\sigma$-model shown
here is that, for
finite values of $\tau/\tau_{\rm tr}$, the expansion of the free
energy may be continued to find further terms that serve as
corrections to the usual form of the ballistic $\sigma$-model,
Eq.~(\ref{M+K}), but may nevertheless become
relevant, as we discuss further below.

Our aim now is to integrate over the $P_{{\bf n}}$ fields by treating them in a
Gaussian approximation. To do so we find it convenient to Fourier transform
from $P_{{\bf n}}({\bf r})$ to angular harmonic and momentum space via 
\[
P_{{\bf n}}({\bf r})=\int \sum_{m}P_{m,{\bf k}}\exp (i({\bf kr}+m\varphi
)\tau _{3})\frac{d^{2}{\bf k}}{(2\pi )^{2}}, 
\]
where $\varphi $ is the polar angle of ${\bf n}$. Performing the harmonic
expansion on the collision integral, we see that, as well as the SCBA
scattering lifetime described previously, a whole series of lifetimes
associated with successive harmonics appears. We define the $m$th lifetime $%
\tau ^{(m)}$ by 
\begin{equation}
\frac{\Lambda }{2\tau ^{(m)}}=2\int \frac{d^{2}{\bf p}_{1}}{(2\pi )^{2}}V_{%
{\bf p}-{\bf p}_{1}}^{ij}p_{1}^{i}p_{1}^{j}\Lambda _{{\bf p}_{1}}\cos
(m\varphi ),  \label{b9}
\end{equation}
where $\varphi $ is the angle between ${\bf p}$ and ${\bf p}_{1}$, so that $%
\tau ^{(0)}$ coincides with the SCBA $\tau $ and $1/\tau _{{\rm tr}}=1/\tau
-1/\tau ^{(1)}$. For example, the collision term in Eq.~(\ref{M+K}) becomes 
\begin{eqnarray}
F_{{\rm coll}} &=&-\frac{\pi \nu }{4}{\rm Str}\int w({\bf n},{\bf n}^{\prime
})Q_{{\bf n}}({\bf r})Q_{{\bf n}^{\prime }}({\bf r})\frac{d{\bf n}d{\bf n}%
^{\prime }}{(2\pi )^{2}}d{\bf r}  \nonumber \\
&&\hspace{-0.8cm}=\pi \nu \int \frac{d^{2}{\bf k}}{(2\pi )^{2}}%
\sum_{m=1}^{\infty }{\rm Str}P_{m,{\bf k}}P_{-m,-{\bf k}}\left( \frac{1}{%
\tau }-\frac{1}{\tau ^{(m)}}\right)  \label{collint}
\end{eqnarray}
\ \ \ \ \ \ \ 

While the form of the free energy given by Eq.~(\ref{M+K}) with the
collision integral (\ref{collint}) is sufficient to treat the case of
long-ranged disorder correlations (for which $\tau _{{\rm tr}}\gg \tau $),
in the limit of short-range disorder certain further terms are also
relevant and should be included for the integration over the nonzero 
harmonics. These extra terms represent 
corrections to the usual form of the ballistic $\sigma $%
-model. These terms are found by continuing the expansion of the logarithm
in Eq.~(\ref{Lag2}) to include terms of order ${\cal A}_{{\rm coll}}^{2}$, $%
{\cal A}_{{\rm kin}}{\cal A}_{{\rm coll}}$, ${\cal A}_{{\rm kin}}{\cal A}_{%
{\rm coll}}^{2}$ and ${\cal A}_{{\rm kin}}^{2}$. In doing so, we ensure that
we include {\em all} 
terms in the free energy that contribute to Gaussian order in 
$P_{{\bf n}}$ and lead to no more than two gradient operators in each term
in the final free energy. Collecting the various terms in the gradient
expansion together, we come to the free energy $F=F_{0}+F_{\shortparallel
}+F_{\perp }+F_{{\rm unit}}$ where %\end{multicols}
%\widetext
\begin{eqnarray}
F_{0} &=&\pi \nu \int \hspace{-0.1cm}\frac{d^{2}{\bf k}}{(2\pi )^{2}}%
\sum_{m=1}^{\infty }{\rm Str}\left[ P_{m,{\bf k}}P_{-m,-{\bf k}}\frac{\tau
^{(m)}}{\tau }\left( \frac{1}{\tau }-\frac{1}{\tau ^{(m)}}\right) \right. 
\nonumber \\
&&\left. +\frac{iv_{F}}{2}\Lambda \left( P_{m,{\bf k}}P_{-m-1,-{\bf k}}\bar{k%
}^{\ast }+P_{m+1,{\bf k}}P_{-m,-{\bf k}}\bar{k}\right) \right] ,  \nonumber
\\
F_{\shortparallel } &=&\pi \nu v_{F}\int d{\bf r}\sum_{m=1}^{\infty }{\rm Str%
}\left[ \Phi _{x}^{\shortparallel }\tau _{3}\Lambda \left(
P_{m}P_{-m-1}+P_{m+1}P_{-m}\right) -i\Phi _{y}^{\shortparallel }\Lambda
(P_{m}P_{-m-1}-P_{m+1}P_{-m})\right] ,  \nonumber \\
F_{\perp } &=&\frac{-i\pi \nu v_{F}}{2}\int d{\bf r}\,{\rm Str}\left[ \Phi
_{x}^{\perp }\tau _{3}\Lambda (P_{1}+P_{-1})+i\Phi _{y}^{\perp }\Lambda
(P_{1}-P_{-1})\right] ,  \nonumber \\
F_{{\rm unit}} &=&\frac{\pi \nu }{8}\int d{\bf r}\,{\rm Str}\left[
D_{0}(\nabla Q)^{2}+2i\omega \Lambda Q\right] ,  \label{Lag3}
\end{eqnarray}
and $\bar{k}=k_{x}+ik_{y}\tau _{3}$. The supermatrices ${\bf \Phi }$ and $Q$
are defined by
\begin{equation}
{\bf \Phi }=\bar{U}{\bf \nabla }U,\text{ \ }Q({\bf r})=U({\bf r})\Lambda 
\bar{U}({\bf r})  \label{Lag3a}
\end{equation}
and $\Phi ^{\shortparallel }$ ($\Phi ^{\perp }$) are the components of ${\bf %
\Phi }$ that commute (anti-commute) with $\Lambda $. The parameter $D_{0}$
is equal to $D_{0}=v_{F}^{2}\tau /2$. As an intermediate step we have
rescaled $P_{m}\rightarrow P_{m}\tau ^{(m)}/\tau $. Eqs.~(\ref{Lag3}) now
represent the free energy that is applicable for energy scales that are much
less than $\tau ^{-1}$, but may still be as large as $\tau _{{\rm tr}}^{-1}$.

In the limit of short-range (delta-correlated) disorder ($\tau^{(m)}\to
\infty$ for $m\neq 0$), we see that the $P_m$ fields in Eq.~(\ref{Lag3}) are
now infinitely massive. This observation requires the
presence of the extra terms that we have included in addition to those of
usual ballistic $\sigma$-model, Eq.~(\ref{M+K}). 
These terms lead to the appearance of the extra 
$(\tau^{(m)}/\tau)$ factor in $F_0$, as compared to expression (\ref{collint}%
). Due to the infinite mass of the $P_m$ fields, only the terms in $F_{{\rm %
unit}}$ remain in the free energy and the conventional unitary $\sigma$%
-model is recovered in this limit, as required.

As a technical point we remark that our form of the free energy (\ref{Lag3})
differs from the ballistic $\sigma $-model as written in Eq.~(\ref{M+K}) in
a further sense: the lifetimes $\tau ^{(m)}$ that appear in our expressions
are defined through a self-consistent Born approximation, according to Eq.~(%
\ref{b9}). This expression allows in general for off-shell contributions
from the momenta ${\bf p}_{1}$. The collision integral in the model of Eq.~(%
\ref{M+K}), by contrast, amounts to only a simple Born approximation (see
Eq.~(\ref{tauw})) as it includes only on-shell contributions through the
factor $w({\bf n},{\bf n}^{\prime })$. For the RMF model with a finite $%
\kappa $, the distinction is not important for sufficiently low angular
harmonics, as then the relevant momentum integrals are restricted to the
energy shell anyway due to the compensation of the singularity in $w({\bf n},%
{\bf n}^{\prime })$ as ${\bf n}\rightarrow {\bf n}^{\prime }$ by the $(Q_{%
{\bf n}}-Q_{{\bf n}^{\prime }})^{2}$ factor in Eq.~(\ref{M+K}). As an
example, for $m=1$ this compensation has been encountered already in Eq.~(%
\ref{tautr}) for $\tau _{{\rm tr}}$. For harmonics $m$ of the order of $%
\kappa ^{-1}$, however, the simple Born approximation becomes unreliable and
instead the full momentum dependence of $V_{{\bf p}-{\bf p}_{1}}$ must be
accounted for. For such high harmonics, the factor of $\tau ^{(m)}/\tau $ in 
$F_{0}$ becomes much larger than unity and hence the contribution of these
harmonics is strongly damped. The end result of these considerations is that
for the RMF model 
we may apply a cutoff to the angular harmonic summation in Eq.~(\ref{Lag3})
to $m\ll \kappa ^{-1}$.

\subsection{Integration over nonzero harmonics}

\label{sec:nonzero} Having derived the free energy functional of Eq.~(\ref
{Lag3}) for fluctuations at energy scales much less than $\tau ^{-1}$, the
next step is to reduce this form, by an integration over the non-zero
harmonics $P_{{\bf n}}({\bf r})$, to one that is applicable at the lowest
energy scales, which are much less than $\tau _{{\rm tr}}^{-1}$. In other
words, we need to average the terms in the free energy that couple the (b)
and (c) modes ($F_{\shortparallel }$ and $F_{\perp }$) with respect to the
bare (b) mode free energy ($F_{0}$) to produce the required renormalization
of the bare (c) mode free energy ($F_{{\rm unit}}$). Equivalently, we aim
to determine the influence of ballistic 
electron motion on distances smaller that $%
l_{{\rm tr}}$, at which distances the classical limit is described
by the Boltzmann equation, on quantum interference processes at large
distances.

Since relevant terms in the free energy will contain no more than two
gradient operators, the relevant contribution from this averaging comes from
the second-order cumulant of $F_{\shortparallel }+F_{\perp }$: 
\begin{equation}
F_{0}+F_{\shortparallel }+F_{\perp }\rightarrow -\frac{1}{2}\langle
(F_{\shortparallel }+F_{\perp })^{2}\rangle _{F_{0}}.  \label{c1}
\end{equation}

While the contribution from the cross term $F_{\shortparallel }F_{\perp }$
vanishes, the terms in $F_{\perp }$ may be eliminated by a simple shift in $%
P_{\pm 1}$ which leads to a dressing of the bare diffusion coefficient, $%
D_{0}$, appearing in $F_{{\rm unit}}$: $D_{0}\rightarrow D=v_{F}^{2}\tau _{%
{\rm tr}}/2$. In terms of perturbation theory, the replacement of the bare $%
D_{0}$ by the classical diffusion coefficient $D$ corresponds as usual to
the inclusion of two-particle vertex corrections. Taking into account only
the contributions from $F_{{\rm unit}}$ and $F_{\perp }$ corresponds to the
calculation of Refs.~\cite{woelfle,Aronov} and gives the conventional
unitary $\sigma $-model with the free energy functional $F_{{\rm unit}}$,
Eq. (\ref{Lag3}), and the classical coefficient $D$.

What remains is to evaluate the contribution from the
terms in $F_{\shortparallel }$.
 In contrast to the
contributions from $F_{\perp }$ and $F_{{\rm unit}}$, which involve only the
zeroth and first harmonics, the contribution from $F_{\shortparallel }$
involves correlations between higher harmonics. To calculate this
contribution we need to perform the set of Gaussian integrals that
correspond to the integration over $P_{m}$ with the free energy functional $%
F_{0}$. This step is no more than an application of Wick's theorem in the
space of the angular harmonics (see also Aleiner and Larkin\cite{Aleiner}
for a further example) and so requires inversion of the quadratic form in $%
P_{m}$ contained in the bare free energy $F_{0}$.

The full form of the Gaussian integration over the $P_{m}$ fields is made
clear if we write out the components of $P_{m}$ as 
\[
P_{m,{\bf k}}=\left( 
\begin{array}{cc}
0 & B_{m,{\bf k}} \\ 
(\bar{B})_{m,{\bf k}} & 0
\end{array}
\right) ,\,B_{m,{\bf k}}=\left( 
\begin{array}{cc}
a_{m,{\bf k}} & i\sigma _{m,{\bf k}} \\ 
\rho _{-m,-{\bf k}}^{\dagger } & ib_{m,{\bf k}}
\end{array}
\right) , 
\]
We see from the form of $F_{0}$ that there are no correlations between
negative and positive harmonics of $B_{m,{\bf k}}$; defining the column
vectors $\vec{a}_{\pm }=(a_{\pm 1},a_{\pm 2},\ldots )$, $\vec{b}_{\pm
}=(b_{\pm 1},b_{\pm 2},\ldots )$ and similarly for $\vec{\rho}$ and $\vec{%
\sigma}$, we come to $F_{0}=F_{0}^{+}+F_{0}^{-}$ where 
\begin{eqnarray}
F_{0}^{\pm } &=&\frac{2\pi \nu }{\tau _{{\rm tr}}}\int \frac{d^{2}{\bf k}}{%
(2\pi )^{2}}[\vec{a}_{\pm ,{\bf k}}^{\,\ast }\hat{L}(\bar{s}_{\pm })\vec{a}%
_{\pm ,{\bf k}}+\vec{b}_{\pm ,{\bf k}}^{\ast }\hat{L}(\bar{s}_{\pm })\vec{b}%
_{\pm ,{\bf k}}  \label{b10} \\
&&+\vec{\sigma}_{\pm ,{\bf k}}^{\ast }\hat{L}(\bar{s}_{\pm })\vec{\sigma}%
_{\pm ,{\bf k}}-\vec{\rho}_{\pm ,{\bf k}}^{\,\ast }\hat{L}(\bar{s}_{\pm })%
\vec{\rho}_{\pm ,{\bf k}}],  \nonumber \\
\bar{s}_{\pm } &=&\frac{l_{{\rm tr}}}{2}(k_{x}\pm i\tau _{3}k_{y}). 
\nonumber
\end{eqnarray}

Here $\hat{L}$ is a tridiagonal, semi-infinite matrix with the entries 
\begin{equation}
\left( \hat{L}(\bar{s})\right) _{m,m^{\prime }}\hspace{-0.1cm}=\frac{\tau _{%
{\rm tr}}\tau ^{\left( m\right) }}{\tau }\left( \frac{1}{\tau }-\frac{1}{%
\tau ^{\left( m\right) }}\right) \delta _{m,m^{\prime }}+i\bar{s}\delta
_{m+1,m^{\prime }}+i\bar{s}^{\ast }\delta _{m,m^{\prime }+1},  \label{Lform}
\end{equation}
where $m$ and $m^{\prime}$ are positive and 
\[
\bar{s}=\frac{l_{{\rm tr}}}{2}\left( k_{x}+ik_{y}\right)
\]

After a similar reexpression of $F_{\shortparallel }$ in terms of the
vectors $\vec{a}$, $\vec{b}$, $\vec{\sigma}$ and $\vec{\rho}$, one can
perform the averaging. In the process we use the relations ${\rm Str}\Phi =%
{\rm Str}(\Lambda \Phi )={\rm Str}(\tau _{3}\Phi )=0$ that follow from the
symmetries of $U$ given by Eq.~(\ref{Usymm}): hence only 
the combination ${\rm Str}%
(\tau _{3}\Lambda \Phi )$ can enter the final formulae. Integrating over the
supermatrices $P_{m}$ in $\left\langle F_{\parallel }^{2}\right\rangle $,
Eq. (\ref{c1}), with the free energy $F_{0}$, Eq. (\ref{Lag3}), we 
obtain a term quadratic in ${\rm Str}(\tau _{3}\Lambda \Phi )$ with
a coefficient determined by an integral over ${\bf k}$. Indeed we 
reduce the
additional term $F_{c}$ in the free energy to the form 
\begin{equation}
F_{c}=-\frac{l_{{\rm tr}}^{2}}{16}\int {\rm Str}\left( \tau _{3}\Lambda \Phi
_{i}\left( {\bf r}\right) \right) {\rm Str}\left( \tau _{3}\Lambda \Phi
_{j}\left( {\bf r}^{\prime }\right) \right) \Pi ^{ij}\left( {\bf r,r}%
^{\prime }\right) d{\bf r}d{\bf r}^{\prime },  \label{b11}
\end{equation}
where 
\begin{equation}
\Pi ^{ij}\left( {\bf r,r}^{\prime }\right) =\int n^{i}n^{\prime j}\Gamma
\left( {\bf r,r}^{\prime };{\bf n,n}^{\prime }\right) \Gamma \left( {\bf r}%
^{\prime },{\bf r;n}^{\prime }{\bf ,n}\right) \frac{d{\bf n}d{\bf n}^{\prime
}}{\left( 2\pi \right) ^{2}}.  \label{b12}
\end{equation}
The function $\Gamma \left( {\bf r,r}^{\prime };{\bf n,n}^{\prime }\right) $
depends only on the coordinate 
difference ${\bf r-r}^{\prime }$ and may be written in
the momentum representation as
\begin{equation}
\Gamma \left( {\bf k;n,n}^{\prime }\right) =\sum_{m,m^{\prime }>0}
\left[\left( 
\hat{L}^{-1}\left( \bar{s}\right) \right) _{m,m^{\prime }}e^{i\left( m\phi
-m^{\prime }\phi ^{\prime }\right) }+\left( \hat{L}^{-1}\left( -\bar{s}%
\right) \right) _{m,m^{\prime }}e^{-i\left( m\phi -m^{\prime }\phi ^{\prime
}\right) }\right],  \label{b12a}
\end{equation}
where $\phi $ and $\phi ^{\prime }$ are the polar angles of the vectors ${\bf %
n}$ and ${\bf n}^{\prime }$. An alternative formulation is that 
$\Gamma \left( {\bf r,r}^{\prime
};{\bf n,n}^{\prime }\right) $ satisfies the following Boltzmann-like
equation: 
\begin{eqnarray}
l_{{\rm tr}}{\bf n\nabla }_{{\bf r}}\Gamma \left( {\bf r,r}^{\prime };{\bf %
n,n}^{\prime }\right) +\int W\left( {\bf n,n}^{\prime \prime }\right) \Gamma
\left( {\bf r,r}^{\prime };{\bf n}^{\prime \prime }{\bf ,n}^{\prime }\right) 
\frac{d{\bf n}^{\prime \prime }}{2\pi }
&=&\delta \left( {\bf r-r}^{\prime
}\right) \delta \left( {\bf n-n}^{\prime }\right),  
\label{b14}
\end{eqnarray}
where $W\left( {\bf n,n}^{\prime }\right) $ is a function of ${\bf nn}%
^{\prime }$, such that its eigenvalues are equal to the diagonal
entries of $\hat{L}$, ie. $\tau _{{\rm tr}}\tau
^{\left( m\right) }\tau ^{-1}\left( \tau ^{-1}-\left( \tau ^{\left( m\right)
}\right) ^{-1}\right) .$
As we want to derive a free energy functional for  
supermatrices ${\bf \Phi ,}$ that vary slowly on the scale of 
$l_{{\rm tr}}$, we assume that the
functions ${\bf \Phi }\left( {\bf r}\right) $ depend more slowly
on ${\bf r}$ than 
$\Pi ^{ij}\left( {\bf r,r}^{\prime }\right) $. 

Using Eqs. (\ref{b14}) one can check without difficulty that the function $%
\Pi ^{ij}\left( {\bf r,r}^{\prime }\right) $, Eq. (\ref{b12}), satisfies a
transversality condition 
\begin{equation}
\nabla _{{\bf r}i}\Pi ^{ij}\left( {\bf r,r}^{\prime }\right) =\nabla _{{\bf r%
}^{\prime }j}\Pi ^{ij}\left( {\bf r,r}^{\prime }\right)=0.  \label{b15}
\end{equation}

As the function $\Pi ^{ij}\left( {\bf r,r}^{\prime }\right) $ depends only
on ${\bf r-r}^{\prime }$ we perform the Fourier transformation in this
variable and using Eq. (\ref{b15}) write it in the form 
\begin{equation}
\Pi ^{ij}\left( {\bf q}\right) =\Pi_0\left(q\right) 
\left( \delta ^{ij}-\frac{%
q^{i}q^{j}}{q^{2}}\right).   \label{b16}
\end{equation}

Using Eqs.~(\ref{b11}), we write the final form of the $%
\sigma $-model as 
\begin{eqnarray}
F[Q] &=&\frac{\pi \nu }{8}\int \,{\rm Str}[D(\nabla Q\left( {\bf r}\right)
)^{2}+2i\omega \Lambda Q\left( {\bf r}\right) ]d{\bf r}   \nonumber\\
&&-\frac{l_{{\rm tr}}^{2}}{16}\int {\rm Str}\left( \tau _{3}\Lambda \Phi
_{i}\left( {\bf r}\right) \right) {\rm Str}\left( \tau _{3}\Lambda \Phi
_{j}\left( {\bf r}^{\prime }\right) \right) \Pi ^{ij}\left( {\bf r,r}%
^{\prime }\right) d{\bf r}d{\bf r}^{\prime }.
\label{b100}
\end{eqnarray}

The invariance of the free energy functional $F$ under the local gauge
transformations means that it can be written in terms of the supermatrix $Q$
only (without ${\bf \Phi }$). This may be done by employing the 
transversal form for $\Pi^{ij}({\bf q})$, as given by Eq.~(\ref{b16}),
in the second term of Eq. (\ref{b100}). Then we obtain
\begin{eqnarray}
F[Q] &=&\frac{\pi \nu }{8}\int {\rm Str}[D(\nabla Q\left( {\bf r}\right)
)^{2}+2i\omega \Lambda Q\left( {\bf r}\right) ]d{\bf r}  \label{b101} \\
&&-\frac{l_{\rm tr}^2}{256}\int \frac{\Pi_0(q)}{q^2}
M\left( {\bf q}\right) M(-{\bf q}) \frac{d^2 q}{(2\pi)^2} ,  \nonumber
\end{eqnarray}
where 
\begin{eqnarray}
M\left( {\bf r}\right) ={\rm Str}\left( \tau _{3}Q\left( {\bf r}\right)
[\nabla _{x}Q\left( {\bf r}\right) ,\nabla _{y}Q\left( {\bf r}\right)
]\right) 
\label{Mdef}
\end{eqnarray}
represents the local topological density\cite{Pruisken}. 
The free energy functional $F[Q]$, Eq.~(\ref{b101}), has a similar
form to that written by Zhang and Arovas\cite{arovas} (although the
latter reference does not contain a proper microscopic justification), up to the
presence of the factor $\Pi_0(q)$. 

What is crucial is the behaviour of 
$\Pi_0(q)$ as $q\to 0$. If $\Pi_0(q)$ were to remain finite 
as $q\to 0$, then the extra term in the free energy, Eq.~(\ref{b101}),
would remain relevant and give rise to logarithmic corrections to the
conductivity. A vanishing value of $\Pi_0(q)$ means that the new term
is not relevant, and gives rise to only higher than logarithmic
corrections to the conductivity. These statements are made clear in
section \ref{sec:pert}, where we show how the quantity $\Pi ^{ij}\left( {\bf r,r}^{\prime
    }\right) $ appears in a direct calculation of the conductivity. 
Moreover we show in section \ref{sec:current} that 
the quantity $\Pi ^{ij}\left( {\bf r,r}^{\prime
    }\right) $ has a direct 
 physical meaning: it is proportional the mesoscopic
 current-current correlation function. Thus the question of the
 potential relevancy of the new term in Eq.~(\ref{b101}) amounts to
 whether or not the Fourier transform of the current-current
 correlator remains nonzero in the limit of $q\to 0$. 

In fact the evaluation of the quantity $\Pi_0(q)$ as $q\to 0$ is a
delicate procedure within the framework of the $\sigma$-model. 
As shown very recently by Gornyi {\em et al}.\cite{Gornyi}, a proper evaluation
requires a careful regularisation procedure that
ensures the preservation of gauge invariance. Their proposal is to
evaluate $\Gamma \left( {\bf r,r}^{\prime
};{\bf n,n}^{\prime }\right) $ by a semiclassical approximation and to 
impose a short-distance cutoff via a minimum path length. Regularising 
in this way, the value of $\Pi_0(q)$ turns out to {\em vanish} as
$q\to 0$. Their proposal was supplemented by a diagrammatic
calculation for the current-current correlator which seems to verify
the use of regularisation procedure;
we refer to Ref.\cite{Gornyi} for further details
(we remark that our earlier
letter\cite{TEfetov} employed a more naive regularisation procedure
which led to incorrect conclusions).

This demonstrates that, within the
scope of our analysis, the new term 
seems to remain irrelevant in the
calculation of corrections to the conductivity that are logarithmic in
frequency. As a result we do not find
evidence for deviation of the localization behaviour
of the delta-correlated random magnetic field model from that of the
conventional unitary ensemble.  As such our conclusions coincide
with those of the original work of
Aronov {\em et al}., although for
rather more subtle reasons than were originally realised in that 
work. 

Even so, certain questions remain which
deserve further study before a definitive conclusion can be drawn for
this problem. In particular, the role of the massive (a) modes,
describing fluctuations of the eigenvalues of $Q$ and the Cooperons,
has been neglected in our analysis and should be considered more
carefully. For example, there is some ambiguity regarding the precise
value of the single-particle lifetime, $\tau$, in the RMF model.
Although the SCBA for the RMF model
yields a value of $\tau^{-1}$ that is divergent 
as $\kappa\to 0$, 
it is likely that the true value of $\tau^{-1}$
that couples to the massive (a) modes should remain finite; in a
similar manner, a gauge-invariant
formulation\cite{EAlt} (beyond the SCBA) has been shown to produce a
finite single-particle inverse lifetime,
at least when the average magnetic field is nonzero. In the formalism
of the field theory, this suggests that an improved saddle-point may
be found which does not contain any divergencies in the limit of
$\kappa\to 0$. Since the
single-particle mean free path determines the value of the shortest
lengthscales down to which the ballistic $\sigma$-model is
applicable for this model, and the evaluation of the new term in the free
energy depends crucially on the regularization procedure at such short 
distances, further study of the role of the massive (a) modes in this
problem seems desirable. 
 
Although the new term in the free energy, Eq.~(\ref{b101}), appears not
to give rise to logarithmic corrections to the conductivity for the
delta-correlated RMF, it is still responsible for corrections that are
higher than logarithmic. Such corrections are known as ``memory
effects'' and have been examined recently by Wilke {\em et
  al}.\cite{Wilke}. The derivation of such corrections is contained in 
the calculation of the conductivity in next section.

We remark that a free energy of the form of Eq.~(\ref{b101}) arises
also for a model of long-range potential disorder, as well as vector
potential, as long as time-reversal symmetry is broken by a constant
magnetic field (in which case, a further topological term associated with
Hall quantization appears). If time-symmetry symmetry is preserved,
then the additional term in Eq.~(\ref{b101}) is absent for trivial
symmetry reasons.

The question remains whether the new term in the free energy,
Eq.~(\ref{b101}), is always irrelevant or if such irrelevancy has been specific
to our original choice of model. In section \ref{sec:superdiff} we
show that, indeed, for other models of disorder such a new term {\em
  can} remain relevant, and give logarithmic corrections to the
conductivity. What is required is disorder of even longer range than has been 
envisaged so far, that leads to a finite value of the current-current
correlator as $q\to 0$. This situation leads to
the interesting possibility of a quantum phase transition at weak disorder. 

%********************************************
%* Perturbation Theory                      *
%********************************************
\section{Perturbation Theory}
\label{sec:pert}
In this section we supplement the derivation of the free energy
in the previous sections by direct perturbative calculations within
the field theory formalism. In section \ref{sec:direct}
we calculate the conductivity. We 
identify the appearance in this calculation of the quantity
$\Pi ^{ij}\left( {\bf r,r}^{\prime
    }\right) $, that appears
in the free energy, Eq.~(\ref{b101}), and show how it leads to higher
order than logarithmic corrections to the conductivity. 
We also calculate in section \ref{sec:current} the mesoscopic 
current-current correlation function, and find it is directly
proportional to 
$\Pi ^{ij}\left( {\bf r,r}^{\prime
    }\right) $, thus providing a physical interpretation for this
quantity.

While we operate within 
the framework of a field theory, we emphasize that the 
calculations of this section are
equivalent to conventional diagrammatics.  We remark also that 
the results for the conductivity of section \ref{sec:direct}
may also be derived via an RG analysis for the diffusion coefficient, 
similar to that of section
\ref{sec:RG}: the
equivalence of the two approaches follows from the Einstein relation between
the conductivity and the diffusion coefficient.

%************************************************
%* Conductivity                                 *
%************************************************
\subsection{Conductivity}
\label{sec:direct} 
In this section, we demonstrate a direct calculation of the
conductivity within the framework of the field theory.
According to standard linear response theory (see e.g.\cite{Isihara,Efetov}%
), the response $R$ is defined by the relation of the current density $J$ to
the oscillating part of an applied vector potential $A_{\omega }$, 
\[
J^{i}(\omega )=\frac{i\omega }{c}R^{ij}(\omega )A_{\omega }^{j}.
\]
Following Ref.~\cite{Efetov}, we write the response as $R^{ij}(\omega )=\int
d{\bf r}^{\prime }R^{ij}({\bf r},{\bf r}^{\prime })$, where 
\begin{equation}
R^{ij}({\bf r},{\bf r}^{\prime })=-\frac{e^{2}}{16\pi }\sum_{\gamma ,\delta
=1}^{4}\langle (1-\tau _{3})_{\gamma \gamma }\hat{\pi}_{{\bf r}^{\prime
}}^{i} \, \psi _{\gamma }^{1}({\bf r}^{\prime })\bar{\psi}_{\gamma }^{1}(%
{\bf r})(1-\tau _{3})_{\delta \delta }\hat{\pi}_{{\bf r} }^{j}\, \psi
_{\delta }^{2}({\bf r})\bar{\psi}_{\delta }^{2}({\bf r}^{\prime })\rangle ,
\label{R0}
\end{equation}
and where the averaging is with respect to the Lagrangian ${\cal L}$ in
Eq.~(\ref{Lag0}). Here 
\begin{equation}
\hat{\pi}_{{\bf r}}^{i}\equiv \left( -i\nabla _{{\bf r}}^{i}-\left(
e/c\right) {\bf A}\right) /m  \label{R01}
\end{equation}
is the velocity operator (${\bf A}$ is the static vector potential). On
averaging the right-hand side of Eq.~(\ref{R0}), we may pair the $\psi $
fields, according to Wick's theorem, in three different ways. This leads to $%
R=R_{1}+R_{2}+R_{3}$ where 
\begin{eqnarray}
R_{1}^{ij}({\bf r},{\bf r}^{\prime }) &=&-\frac{e^{2}}{16\pi }{\rm Str}\left[
k(1-\tau _{3})\hat{\pi}_{{\bf r}^{\prime }}^{i}g^{11}({\bf r}^{\prime },{\bf %
r})\right] {\rm Str}\left[ k(1-\tau _{3})\hat{\pi}_{{\bf r}}^{j}g^{22}({\bf r%
},{\bf r}^{\prime })\right] ,  \nonumber \\
R_{2}^{ij}({\bf r},{\bf r}^{\prime }) &=&\frac{e^{2}}{16\pi }{\rm Str}\left[
k(1-\tau _{3})\hat{\pi}_{{\bf r}^{\prime }}^{i}g^{12}({\bf r}^{\prime },{\bf %
r}^{\prime })k(1-\tau _{3})\hat{\pi}_{{\bf r}}^{j}g^{21}({\bf r},{\bf r})%
\right] ,  \nonumber \\
R_{3}^{ij}({\bf r},{\bf r}^{\prime }) &=&\frac{e^{2}}{16\pi }{\rm Str}\left[
k(1-\tau _{3})\hat{\pi}_{{\bf r}^{\prime }}^{i}g^{12}({\bf r}^{\prime },{\bf %
r})\overleftarrow{\pi }_{{\bf r}}^{j}{}k(1+\tau _{3})g^{21}({\bf r}^{\prime
},{\bf r})\right] ,  \label{Reqs}
\end{eqnarray}
and where the symbol $\stackrel{\leftarrow }{\pi }$ indicates a velocity
operator whose derivative acts to the left. The Green function is defined by 
$g^{\alpha \beta }\left( {\bf r},{\bf r}^{\prime }\right) =2\langle \psi
^{\alpha }({\bf r})\bar{\psi}^{\beta }({\bf r}^{\prime })\rangle $ and,
following the introduction of the $Q$-matrix, satisfies (c.f. with Eq. (\ref
{saddle})) 
\begin{equation}
\left( \epsilon +\epsilon _{F}+\frac{\nabla _{{\bf r}}^{2}}{2m}-\frac{\omega 
}{2}\Lambda \right) g({\bf r},{\bf r}^{\prime })-i\int d{\bf r}^{\prime
\prime }\widehat{V}_{{\bf r},{\bf r}^{\prime \prime }}\widetilde{Q}({\bf r},%
{\bf r}^{\prime \prime })g({\bf r}^{\prime \prime },{\bf r}^{\prime
})=-i\delta ({\bf r}-{\bf r}^{\prime }),  \label{green}
\end{equation}
where the function $\hat{V}_{{\bf r,r}^{\prime }\text{ }}$ is determined by
Eq. (\ref{Lint}). In the limit of short-range disorder, $\tau =\tau _{{\rm tr%
}}$, the fluctuations of the nonzero harmonics of $Q_{{\bf n}}({\bf r})$ are
strongly suppressed, as demonstrated in the derivation of the free energy in
section \ref{sec:gradient}. In this case, $Q({\bf r},{\bf r}^{\prime })$
becomes a function of ${\bf R}=({\bf r}+{\bf r}^{\prime })/2$ only and Eq.~(%
\ref{Reqs}) reduces to the relations already found in the direct computation
of the conductivity for short-range disorder (see Ref.~\cite{Efetov},
Chapter 8, and Ref.~\cite{Lerner}). However, in the limit of long range
disorder, when non-zero harmonics are not suppressed, an additional term
appears that contains the quantity $\Pi ^{ij}\left( {\bf r,r}^{\prime
    }\right) $.

In the limit of long-ranged disorder, for which $\tau _{{\rm tr}}\gg \tau $,
we may perform an expansion of the Green function $g({\bf p},{\bf R})$ in
the collision term $\sum_{{\bf p}_{1}}V_{{\bf p}-{\bf p}%
_{1}}^{ij}p_{1}^{i}p_{1}^{j}(Q_{{\bf p}_{1}}-Q_{{\bf p}})$, where the
momentum ${\bf p}$ arises from a Fourier transform with respect to ${\bf \ r}%
-{\bf r}^{\prime }$: this expansion corresponds directly to the expansion of
the free energy in section \ref{sec:gradient} in the quantity ${\cal \ A}_{%
{\rm coll}}$, and corresponds to an expansion in powers of $\tau /\tau _{%
{\rm tr}}\ll 1$. Therefore to leading order in $\tau /\tau _{{\rm tr}}$, we
may take 
\[
g({\bf p},{\bf R})=i\left[ \frac{p^{2}}{2m}-\epsilon +\frac{i}{2\tau }Q_{%
{\bf n}}\right] ^{-1}+\ldots .
\]
Since we neglect the hard massive (a) modes (see section \ref{sec:flucts}),
we have also that $[g,\tau _{3}]=0$ and hence $R_{3}=0$. Performing the
necessary integrals over the modulus of the momentum, we find 
\begin{eqnarray}
R_{1}^{ij} &=&2e^{2}\nu v_{F}^{2}\tau \int \frac{d{\bf n}}{2\pi }n^{i}n^{j}%
\left[ 1-\frac{1}{32}\langle {\rm Str}\left( k(1-\tau _{3})(Q_{{\bf n}}({\bf %
r})-\Lambda )^{11}\right) \times \right.   \nonumber \\
&&\left. \hspace{4cm}{\rm Str}\left( k(1-\tau _{3})(Q_{{\bf n}}({\bf r}%
)-\Lambda )^{22}\right) \rangle \right] ,  \nonumber \\
R_{2}^{ij} &=&\pi \frac{(e\nu v_{F})^{2}}{16}\int d{\bf r}^{\prime }\frac{d%
{\bf n}d{\bf n}^{\prime }}{(2\pi )^{2}}n^{i}n^{\prime j}\,\langle {\rm Str}%
\left[ k(1-\tau _{3})\left[ Q_{{\bf n}}({\bf r})(1+v_{F}\tau {\bf n}\nabla _{%
{\bf r}}Q_{{\bf n}}({\bf r}))\right] ^{12}\right.   \nonumber \\
&&\left. \times k(1-\tau _{3})\left[ Q_{{\bf n}^{\prime }}({\bf r}^{\prime
})(1+v_{F}\tau {\bf n}^{\prime }\nabla _{{\bf r}^{\prime }}Q_{{\bf n}%
^{\prime }}({\bf r}^{\prime }))\right] ^{21}\right] \rangle ,  \label{Rball}
\end{eqnarray}
where now the averaging is with respect to the free energy (\ref{Lag3}). In
the limit of $\tau \ll \tau _{{\rm tr}}$, to leading order Eq.~(\ref{Rball})
reduces to include only the following term $\tilde{R}$, originating from $%
R_{2}$: 
\begin{equation}
\tilde{R}^{ij}=\pi \frac{(e\nu v_{F})^{2}}{16}\int d{\bf r}^{\prime }\frac{d%
{\bf n}d{\bf n}^{\prime }}{(2\pi )^{2}}n^{i}n^{\prime j}\,\langle {\rm Str}%
\left[ k(1-\tau _{3})Q_{{\bf n}}^{12}({\bf r})k(1-\tau _{3})Q_{{\bf n}%
^{\prime }}^{21}({\bf r}^{\prime })\right] \rangle .  \label{Rball2}
\end{equation}
The response $\tilde{R}^{ij}$, Eq.~(\ref{Rball2}), is the contribution that
was absent in the standard treatment of models with short range disorder 
\cite{Lerner,Efetov}. It involves integration over non-zero harmonics and
may be computed by averaging with respect to the free energy (\ref{Lag3}).

As is consistent with the derivation of the final form of the free energy in
section \ref{sec:nonzero}, we employ the parametrization of Eqs.~(\ref{param}%
) and (\ref{a6}), and expand each of the $Q_{{\bf n}}$ matrices in Eq.~(%
\ref{Rball2}) up to second order in $P_{{\bf n}}$. To reproduce the
leading-order quantum correction to the conductivity, we also expand in
the deviation of the $U$ matrix from unity, although we delay this step for
convenience. In other words, 
averaging over $P_{{\bf n}}\left( {\bf r}\right) $ we
derive as the first step an expression for the conductivity in a form of an
integral over supermatrices $Q\left( {\bf r}\right) $ depending on ${\bf r}$
only. Then, one can compute the integral and obtain the final formulae.

To zeroth order in $P_{{\bf n}}$, the expression (\ref{Rball2}) for $R$
vanishes due to the integrations over ${\bf n}$ and ${\bf \ n}^{\prime }$.
On expanding both of the $Q_{{\bf n}}$ matrices in Eq.~(\ref{Rball2}) to
first order in $P_{{\bf n}}$, averaging over $P_{{\bf n}}$, and adding the
contribution of zero order in $P_{{\bf n}}$ from Eq. (\ref{Rball}) we find
the following contributions to $R$: 
\begin{eqnarray}
R_{a}^{ij} &=&2e^{2}\nu D\delta ^{ij}\left[ 1-\frac{1}{32}\langle {\rm Str}%
(k(1-\tau _{3})(Q-\Lambda )^{11}({\bf r}))\times \right.  \nonumber \\
&&\left. \hspace{4cm}{\rm Str}(k(1-\tau _{3})(Q-\Lambda )^{22}({\bf r}%
))\rangle \right] ,  \label{Raeq} \\
R_{b}^{ij} &=&\frac{\pi }{16}e^{2}\nu ^{2}D^{2}\delta ^{ij}\int d{\bf r}%
^{\prime }\langle {\rm Str}\left[ k(1-\tau _{3})(Q\nabla _{{\bf r}^{\prime
}}Q)^{21}({\bf r}^{\prime })\times \right.  \nonumber \\
&&\left. \hspace{6cm}k(1-\tau _{3})(Q\nabla _{{\bf r}}Q)^{12}({\bf r})\right]
\rangle ,  \label{Rbeq}
\end{eqnarray}
where $Q\left( {\bf r}\right) =U\left( {\bf r}\right) \Lambda \bar{U}\left( 
{\bf r}\right) $ and the averaging is now with respect to the final form of
the free energy, Eq.~(\ref{b100}). We remark that to derive the contribution 
$R_{b}$ it is necessary to include a contraction with the terms in the free
energy, Eq.~(\ref{Lag3}), that are linear in $P_{{\bf n}}$.

In the derivation of the contributions $R_{a}$ and $R_{b}$ of Eqs.~(\ref
{Raeq}) and (\ref{Rbeq}), the coupling of only the first harmonics, $P_{\pm
1}$, to the $U$ matrix is involved. As was found in the derivation of the
free energy in section \ref{sec:nonzero}, the role of the first harmonics,
beyond that of the higher harmonics, is to dress the bare diffusion
coefficient, $D_{0}=v_{F}^{2}\tau /2$, with classical vertex corrections, so
that it is renormalized according to $D_{0}\rightarrow D=v_{F}^{2}\tau _{%
{\rm tr}}/2$. For this reason we see that the expressions for $R_{a}$ and $%
R_{b}$ include a renormalized diffusion coefficient, $D$. Moreover, up to
this renormalization, they coincide precisely with the corresponding
expressions that are found\cite{Efetov,Lerner} for the conductivity in the
case of short-range disorder.

An additional contribution, $R_{c}$, comes from expanding both of the $Q_{%
{\bf n}}$ matrices in Eq.~(\ref{Rball2}) to second order in $P_{{\bf n}}$.
On averaging over $P_{{\bf n}}$ with the free energy functional (\ref{Lag3}%
), we find the new contribution that can be written in the form 
\begin{equation}
R_{c}^{ij}=\frac{1}{32\pi }\left( el_{{\rm tr}}\right) ^{2}\int \Pi
^{ij}\left( {\bf r,r}^{\prime }\right) \langle {\rm Str}\left[ k(1-\tau
_{3})Q^{12}({\bf r})k(1-\tau _{3})Q^{21}({\bf r}^{\prime })\right] \rangle d%
{\bf r}^{\prime },  \label{Rceq}
\end{equation}
where the function $\Pi^{ij}$ is given by Eq. (\ref{b12}).
In
contrast to the contributions $R_{a}$ and $R_{b}$, the term $R_{c}$ does not
have a counterpart in the calculation for short-range disorder, since its
presence relies on the long-range disorder correlations that pertain to the
RMF model. In contrast to the derivation of the terms $R_{a}$ and $R_{b}$,
the derivation of $R_{c}$ requires consideration of all harmonics, rather
than only the zeroth and first harmonics.

The final step to compute the quantum correction to the conductivity is to
expand $U$ around unity, according to, for example, 
\begin{equation}
U=\left( \frac{1-iP}{1+iP}\right) ^{1/2},  \label{Uparam}
\end{equation}
in similarity with Eq.~(\ref{c3}). In the same time, one should also
expand in $P$ the free energy, Eq. (\ref{b100}). It is important that the
new term $F_{c}$ leads to terms starting from $P^{4}$ and may
contribute to the conductivity only in higher orders than considered here.
Expanding $R_{a}$, $R_{b}$ and $R_{c}$ to fourth, sixth and second order in $%
P$ respectively, and averaging over $P$, we obtain the conductivity 
\begin{equation}
\sigma ^{ij}=\sigma _{0}\left[\delta ^{ij }( 1+\Delta
_{a}+\Delta _{b})+\Delta _{c}^{ij}\right] ,  \label{sigcor}
\end{equation}
where 
\begin{eqnarray*}
\Delta _{a}&=&\frac{t^{2}}{128}I_{d}^{2},\qquad \Delta _{b}=-\frac{t^{2}}{64d}%
I_{d}^{2},
\end{eqnarray*}
\vspace{-0.5cm}
\begin{eqnarray*}
\Delta _{c}^{ij}&=&\frac{t^{2}l_{\rm tr}^2}{128}
\int \frac{\Pi^{ij}({\bf q})}{(q^{2}+\lambda ^{2})}\frac{d^{d}{\bf q}}
{(2\pi )^{d}},
\end{eqnarray*}
and
\begin{equation}
I_{d}=\int \frac{1}{(q^{2}+\lambda ^{2})}\frac{d^{d}{\bf q}}{(2\pi )^{d}}.
\label{Ikdef}
\end{equation}
Here $\sigma _{0}=2e^{2}\nu D$ is the classical conductivity and $t=8/(\pi
\nu D)$. The contributions $%
\Delta _{a-c}$ originate from the expressions $R_{a-c}$ respectively.

The contributions $\Delta_a$ and $\Delta_b$ correspond to the usual
weak localization corrections to the conductivity for short-range
disorder in the unitary ensemble. Since $\Pi^{ij}({\bf q})\to 0$ as
${\bf q}\to 0$ (see Ref.\cite{Gornyi}), the contribution $R_c$ is of higher-order than
logarithmic. In fact, since $\Pi^{ij}({\bf q})\propto q^2$, $R_c$ 
corresponds to a ``memory effect'' of the
type studied recently in Ref.~\cite{Wilke}, and gives rise to a
nonanalytic correction $\propto |\omega|$ to the conductivity
as $\omega\to 0$. The contributions $%
\Delta _{a-c}$ may all be derived via an RG approach for the diffusion 
coefficient (see section \ref{sec:RG}), as follows by the Einstein
relation.
 
While we have used the $\sigma $-model approach to compute all of the above
diagrams, we emphasize that in principle it is possible to evaluate all of
the above diagrams directly using the bare Green functions and without
reference to an effective Lagrangian. Such a calculation would be
useful as a check for the regularisation procedure used for the
ballistic $\sigma$-model. The calculation
is not simple  as, for example, it becomes necessary to take into account a
proper dressing of the Hikami boxes and vertices with impurity lines \cite
{hikami}. In addition extra, mutually cancelling diagrams would appear that
correspond to the hard massive modes of the field theory. The
diagrammatic  arguments of Gornyi 
{\em et al}.\cite{Gornyi} do not fulfill the purpose of such a check:
although their calculation of the current-current correlation function 
seems valid (see the next subsection), 
their calculation of the conductivity relies on a
relation between the conductivity and the current-current correlation
function which contains the quantity $\tau$. As we mentioned above,
this quantity formally vanishes within the SCBA for the
delta-correlated RMF model. Thus their arguments are 
questionable for the delta-correlated RMF model, 
although they may be applicable when
the RMF is a weak perturbation to sufficiently strong, short-range
disorder, in which case a well-defined $\tau$ does exist. Instead, 
what is required is a
direct diagrammatic calculation of the first logarithmic correction to the
conductivity, a task which is not simple even for short-range
disorder and has not been attempted so far for the RMF model. 

%**********************************************************
%* Current-Current Correlation function                   *
%**********************************************************

\subsection{Mesoscopic Current Correlations}
\label{sec:current}
In this section we continue the perturbative analysis to 
a calculation of the current-current correlation
function for long-ranged disorder, such as a
delta-correlated RMF. This enables us to provide a physical
interpretation of the quantity 
$\Pi ^{ij}\left( {\bf r,r}^{\prime
    }\right) $,
as being directly proportional to the current-current correlation
function. 

We employ the definition
\begin{equation}
J^{i}({\bf r})=\frac{ie}{4\pi}
\lim_{{\bf r}^{\prime }\rightarrow {\bf r}}(\hat{%
\pi}_{{\bf r}}^{i}-\hat{\pi}_{{\bf r}^{\prime }}^{i})(G_{\varepsilon }^{R}(%
{\bf r},{\bf r}^{\prime })-G_{\varepsilon }^{A}({\bf r},{\bf r}^{\prime })),
\label{Jdef}
\end{equation}
for the local current, 
where $\hat{\pi}_{{\bf r}}^{i}$ is the velocity operator,
Eq. (\ref{R01}).  The current-current correlator 
\begin{equation}
I^{ij}({\bf r}-{\bf r}^{\prime })\equiv \langle J^{i}({\bf r})J^{j}({\bf r}%
^{\prime })\rangle _{{\rm imp}}  \label{meso1}
\end{equation}
is then given by 
\begin{eqnarray}
I^{ij}({\bf r}-{\bf r}^{\prime }) &=&-\frac{e^{2}}{32\pi^2}\lim_{{\bf s}%
\rightarrow {\bf r},{\bf s}^{\prime }\rightarrow {\bf r}^{\prime }}\langle
(1-\tau _{3})_{\gamma \gamma }(\pi _{{\bf r}}^{i}-\pi _{{\bf s}}^{j})\psi
_{\gamma }^{1}({\bf r})\bar{\psi}_{\gamma }^{1}({\bf s})\times 
\label{meso1a} \\
&&\hspace{2cm}\times (1-\tau _{3})_{\delta \delta }(\pi _{{\bf r}^{\prime
}}^{i}-\pi _{{\bf s}^{\prime }}^{j})\psi _{\delta }^{2}({\bf r}^{\prime })%
\bar{\psi}_{\delta }^{2}({\bf s}^{\prime })\rangle _{{\cal L}}  \nonumber
\end{eqnarray}
where a summation over $\gamma $ and $\delta $ is implied.
The angle brackets in Eq. (\ref{meso1}) stand for 
averaging over impurities, while the angle brackets in
Eq. (\ref{meso1a}) stand for averaging with the Lagrangian $%
{\cal L}$, Eq. (\ref{Lag0}).

There are now three possible ways of making pairings of the $\psi $ and $%
\bar{\psi}$ fields. Note that the calculation is following very similar
lines to those of the calculation of the conductivity in section \ref
{sec:direct}. As in that case, only one of the pairings need be considered
at leading order, where the first $\psi $ and $\bar{\psi}$ are paired, and
the second $\psi $ and $\bar{\psi}$ are paired. Following Fourier
transformations with respect to ${\bf r}-{\bf s}$ and ${\bf r}^{\prime }-%
{\bf \ s}^{\prime }$, we find 
\begin{eqnarray}
I^{ij}({\bf r}-{\bf r}^{\prime }) &=&-\frac{e^{2}}{128\pi^2}\lim_{{\bf s}%
\rightarrow {\bf r},{\bf s}^{\prime }\rightarrow {\bf r}^{\prime }}{\rm Str}%
(k(1-\tau _{3})(\pi _{{\bf r}}^{i}-\pi _{{\bf s}}^{i})g^{11}({\bf r},{\bf s}%
))\times   \label{meso2} \\
&&\hspace{2cm}\times {\rm Str}(k(1-\tau _{3})(\pi _{{\bf r}^{\prime
}}^{j}-\pi _{{\bf s}^{\prime }}^{j})g^{22}({\bf r}^{\prime },{\bf s}^{\prime
})),  \nonumber
\end{eqnarray}
where $g({\bf r},{\bf s})$ is defined as in section \ref{sec:direct}, Eq. (%
\ref{green}). Following Fourier transformations with respect to ${\bf r}-%
{\bf s}$ and ${\bf r}^{\prime }-{\bf \ s}^{\prime }$, integrating over the
modulus of the momentum, and keeping the leading-order term in the limit of $%
\tau \ll \tau _{{\rm tr}}$, we arrive at 
\begin{eqnarray}
I^{ij}({\bf r}-{\bf r}^{\prime }) &=&-\frac{e^{2}}{128}(\nu
v_{F})^{2}\int \frac{d{\bf n}d{\bf n}^{\prime }}{(2\pi )^{2}}n^{i}n^{\prime
}{}^{j}{\rm Str}(k(1-\tau _{3})Q_{{\bf n}}({\bf r})(1+\Lambda ))\times  
\nonumber \\
&&\hspace{3cm}\times {\rm Str}(k(1-\tau _{3})Q_{{\bf n}^{\prime }}({\bf r}%
^{\prime })(1-\Lambda )).  \label{ccexp}
\end{eqnarray}
Eq.~(\ref{ccexp}) represents the required formula from which the
current-current correlation function may be computed (compare with Eq.~(\ref
{Rball2}) for the conductivity). As for the conductivity, we employ the
parametrization of Eqs.~(\ref{param}) and (\ref{a6}), and expand each of
the $Q_{{\bf n}}$ matrices in Eq.~(\ref{ccexp}) up to second-order in $P_{n}$%
. If we expand both of the $Q_{%
{\bf n}}$ terms in Eq.~(\ref{ccexp})
to second-order in $P_{{\bf n}}$, we find the contribution
\begin{eqnarray}
I_0^{ij}({\bf r}-{\bf r}^{\prime }) &=&\frac{e^{2}}{32}(\nu
v_{F})^{2}\int \frac{d{\bf n}d{\bf n}^{\prime }}{(2\pi )^{2}}n^{i}n^{\prime
}{}^{j}{\rm Str}(k(1-\tau _{3})P_{{\bf n}}({\bf r})^2(1+\Lambda ))\times  
\nonumber \\
&&\hspace{3cm}\times {\rm Str}(k(1-\tau _{3})P_{{\bf n}^{\prime }}({\bf r}%
^{\prime })^2(1-\Lambda )).  \label{cc2}
\end{eqnarray}
Performing the
contractions in Eq.~(\ref{cc2}), we find
\begin{equation}
I_0^{ij}({\bf r}-{\bf r}^{\prime })=\frac{e^2}{2\pi^2}
l_{{\rm tr}}^{2}\Pi ^{ij}\left( 
{\bf r,r}^{\prime }\right)   \label{meso3}
\end{equation}
where the function $\Pi ^{ij}\left( {\bf r,r}^{\prime }\right) $ is given by
Eq. (\ref{b12}). 
We see from Eq. (\ref{meso3}) that the correlator of mesoscopic currents $I_0^{ij}$
depends on the same function $\Pi ^{ij} $
that enters the action, Eq. (\ref{b11}), 
and the conductivity, Eq.~(\ref{Rceq}%
). 

Strictly speaking, the function $\Pi^{ij}$ that appears in the
free energy represents not
precisely the current-current correlation function, but the
contribution of only the nonzero harmonics to this quantity: this is
reflected by the fact that $I^{ij}$ contains further contributions,
beyond $I_0^{ij}$, 
which are zeroth order in $P_{\bf n}$, but second-order in the
generator $P$ of $U$ (see the parametrization of
Eq.~(\ref{Uparam})). In a similar way, the function 
$\Gamma \left( {\bf k};{\bf n,n}^{\prime }\right) $ of
Section \ref{sec:nonzero} is
defined in Eq.~(\ref{b12a}) through a projection onto only 
nonzero harmonics $m>0$. An alternative definition of 
$\Gamma \left( {\bf k};{\bf n,n}^{\prime }\right) $ may be 
written which treats the zeroth and nonzero harmonics on an equal
footing, as in
Ref.~\cite{Gornyi}, although the difference becomes inessential 
for momenta $k \gg l_{\rm tr}^{-1}$; the deviation 
of the function $\Pi^{ij}$ from the true value of the current-current
correlation function is therefore minor. The separation of
zeroth and nonzero harmonics is in fact inessential in the computation of
correlation functions, as performed in this section and in Ref.~\cite{Gornyi};
however, it is unavoidable for the derivation of a free energy in
terms of a local $Q({\bf r})$ matrix that represents only the
zeroth harmonics, as has been performed in section \ref{sec:functional}.

%****************************************************************
%* Classical Superdiffusion                                     *
%****************************************************************

\section{Classical Superdiffusion}

\label{sec:superdiff} 
We now generalise the discussion to include the possibility of
longer-ranged disorder than has been represented so far by the
delta-correlated RMF. The question that arises from the preceding
analysis is whether a new term, of the type represented in the free energy of 
Eq.~(\ref{b101}), may exist and still remain relevant to give 
new, logarithmic corrections to the conductivity. As has been
demonstrated in section \ref{sec:nonzero}, the potential relevancy of
such a term is related to the question of whether the current-current 
correlator
remains nonzero in the limit of $q\to 0$. Although such a situation does
not arise for the model of a delta-correlated RMF, it is not ruled out 
for models of longer ranged disorder. 

In fact, the problem of diffusion in a field of currents that are
correlated in precisely such a way, ie. so that the current-current correlator
remains nonzero as $q\to 0$, has already been examined on a classical
level\cite
{Derrida,KravLern,Fisher}. These authors considered a model of
classical $2D$ diffusion in a random stationary velocity field, as
governed by a classical Fokker-Planck equation 
\begin{equation}
\left[ \partial /\partial t+{\bf \nabla }\left( {\bf v-}D_{0}{\bf \nabla }%
\right) \right] P\left( {\bf r,}t\right) =0.  \label{dif1}
\end{equation}
Here $P\left( {\bf r},t\right) $ is the distribution function of a randomly
walking particle, where $D_{0}$ is the classical diffusion coefficient. The
randomness of the velocity ${\bf v}$ makes the problem in $2D$ quite
non-trivial. One can imagine different forms of the
velocity-velocity correlations and so, different models leading to
different types of diffusive behavior were considered. 
In one of the models (Model II
in classification of Ref. \cite{KravLern}) the following correlation was
assumed: 
\begin{equation}
\langle v^{i}({\bf r})v^{j}({\bf r}^{\prime })\rangle =\gamma _{0}K^{ij}(%
{\bf r}-{\bf r}^{\prime }),  \label{dif2}
\end{equation}
where the function $K^{ij}\left( {\bf r-r}^{\prime }\right) $ is 
defined as
\begin{eqnarray}
K^{ij}\left( {\bf r-r}^{\prime }\right) &=&\delta ^{ij}\delta \left( {\bf r-r}%
^{\prime }\right) -\frac{1}{2\pi }\nabla _{{\bf r}}^{i}\nabla _{{\bf r}%
^{\prime }}^{j}\ln |{\bf r-r}^{\prime }|, 
\label{b182}
\end{eqnarray}
so that in momentum space, it takes the form
\begin{equation}
K^{ij}\left( {\bf q}\right) =\delta ^{ij}-\frac{q^{i}q^{j}}{q^{2}}.
\label{dif2a}
\end{equation}
The model of Eqs.~(\ref{dif1}) and (\ref{dif2}) 
may describe, for example, Brownian motion of a particle in an
incompressible liquid with random stationary flow. 
Although roughly speaking the function 
$K^{ij}$ describes short-range correlations of the liquid, more
precisely long ranged correlations are present 
in Eq.~(\ref{dif2}) due to the incompressibility. 

In Refs.~\cite{KravLern,Fisher} a Green function of Eq. (\ref{dif1}) was
introduced and expressed as a
functional integral over a vector field ${\bf \phi }$ (in
Ref.~\cite{KravLern} a replica formulation was used).
In this way calculation of the diffusion coefficient $D$ is reduced to 
a study of $\phi ^{4}$ theory. In $2D$ the corrections to the bare diffusion
coefficient are logarithmic, and hence 
a renormalization group treatment may be employed, 
up to the two loop approximation in Ref.~\cite{KravLern}. The
dependence of the diffusion coefficient $D$ on frequency $\omega $ was
obtained in the form 
\begin{equation}
D\left( \omega \right) =D_{0}\left[ 1+g_{0}\ln \left(
\frac{D_{0}k_{0}^{2}}{\omega 
}\right)\right] ^{1/2}  \label{dif3}
\end{equation}
where $k_{0}$ is an ultraviolet cutoff and $g_{0}=\gamma _{0}/\left( 4\pi
D_{0}^{2}\right) $. The dependence $D\left( \omega \right) $ given by 
Eq. (\ref{dif3}) is classified
as superdiffusion because the diffusion coefficient grows as the frequency $%
\omega $ vanishes.

With this background in mind, we return to the problem of electron
motion in the presence of long-ranged disorder. To make closer contact 
with the diffusion-advection problem, we in fact
consider electron motion in the presence of both short range
and long range disorder. We assume that at short distances the
electron motion is determined by the short range potential and is diffusive.
The long range potential, which may be scalar or vector potential or a 
combination of both, influences the diffusion at large distances and may
be considered as a random force.  

Remarkably, we find a direct relation of this electron
model with the
diffusion-advection model, described by Eq.~(\ref{dif1}), as far as
certain terms in the 
solution for the diffusion coefficient are concerned 
(see Eq.~(\ref{dif3})). In the electron model, we find two different
contributions, of opposite sign, in the solution for the diffusion
coefficient. The  positive part of the contribution corresponds
precisely to the term appearing in Eq.~(\ref{dif3}) and hence
corresponds to classical superdiffusion. It
derives from a new term in the free energy, which is of the precisely
the form written in Eq.~(\ref{b101}), although with a finite value of the
current-current correlator ($\Pi_0(q)$) as $q\to 0$; the term
therefore is of the same form as that written originally by Zhang and
Arovas\cite{arovas}. The electron model differs however from the
diffusion-advection model in that it is quantum mechanical:
hence an additional negative contribution appears in the solution for
the diffusion coefficient. This negative contribution is the 
usual unitary weak localization correction due to quantum
interference, and competes
with the above classical correction. This competition leads to the
interesting possibility of a quantum phase transition at weak
disorder, tuned by the relative disorder strengths.

In order to derive a field theory for the electron model we
apply the formalism developed in the preceding sections. We 
proceed by averaging first over the short-range
impurities, but not over the long range potential. A local, quartic term
appears in the Lagrangian in the usual way, which may be decoupled with a
local $Q({\bf r})$ matrix. We find the Lagrangian 
\begin{eqnarray}
{\cal L}&=&\int 
\left[ -\frac{1}{2}{\rm Str}\ln \left( \frac{1}{2m}\left( {\bf %
p}-\frac{e}{c}{\bf A}({\bf r})-i\tau _{3}\nabla _{{\bf r}}\right)
^{2}+V\left( {\bf r}\right) -\epsilon _{F}+\frac{\omega\Lambda}{2}
+\frac{i}{2\tau }Q({\bf r})\right)
\right. \nonumber\\
&&\left.
\hspace{9cm}
+\frac{\pi \nu }{8\tau }{\rm Str}Q^{2}({\bf r})\right] d{\bf r},
\label{dif4}
\end{eqnarray}
where $\tau $ is the mean free time for scattering from the short-range
impurities (corresponding to $D_{0}$ in Eq. (\ref{dif1})). The vector
potential ${\bf A}\left( {\bf r}\right) $ and the long range potential $%
V\left( {\bf r}\right) $, both of which may be random, are not specified now.

Next we write $Q\left( {\bf r}\right) =U\left( {\bf r}\right) \Lambda \bar{U}%
\left( {\bf r}\right) $, cycle the $U$ and $\bar{U}$ factors under the
supertrace and perform a gradient expansion in $\bar{U}\nabla U\tau _{3}$.
To first order in the expansion, we find the term 
\begin{equation}
\delta F=\frac{1}{2m}\int d{\bf r}\frac{d^2 {\bf p}}
{(2\pi )^{2}}{\rm Str}\left( 
\bar{U}\nabla U({\bf r})\tau _{3}\hat{g}_{{\bf p}}({\bf r})\left( {\bf p}-%
\frac{e}{c}{\bf A}\right) \right) ,  \label{Feq1}
\end{equation}
where the matrix function $\hat{g}_{{\bf p}}$ is defined as 
\[
\left[ \frac{1}{2m}\left( {\bf p}-\frac{e}{c}{\bf A}({\bf r})-i\tau
_{3}\nabla _{{\bf r}}\right) ^{2}+V\left( {\bf r}\right) -\epsilon _{F}+%
\frac{i\Lambda }{2\tau }\right] \hat{g}_{{\bf p}}({\bf r})=i.
\]
Note that in the absence of the long range potentials, 
the term $\delta F$ would vanish by symmetry on integration over ${\bf p}$%
. We notice further that the term ${\delta F}$
may be expressed in terms of the local classical current density $%
{\bf J}$, which has been defined in Eq.~(\ref{Jdef}).
Following standard transformations, ${\bf J}$ may be reexpressed in terms of $%
\hat{g}$ as 
\[
J^{i}({\bf r})=\frac{e}{2\pi m}%
\int \frac{d^2{\bf p}}{(2\pi )^{2}}\left( p^{i}-\frac{e}{c}A^{i}({\bf r}%
)\right) (\hat{g}_{{\bf p}}^{11}({\bf r})
-\hat{g}_{{\bf p}}^{22}({\bf r}))_{\alpha\alpha},
\]
where there is no summation over $\alpha$,
which allows us to rewrite Eq.~(\ref{Feq1})
in the form 
\begin{equation}
\delta F=\frac{\pi}{2e}\int d{\bf r}J^{i}({\bf r}){\rm Str}(\Lambda \tau
_{3}\Phi ^{i}\left( {\bf r}\right) ).  \label{dif5}
\end{equation}
where ${\bf \Phi}({\bf r}) 
=\bar{U}\left( {\bf r}\right) \nabla U({\bf r})$. Eq. (%
\ref{dif5}) clarifies immediately the physical meaning of the supermatrix $%
{\bf \Phi} ^{\parallel }$, 
showing that it plays the role of an effective vector
potential. If the particle number is conserved, the current is transversal ($%
{\rm div}{\bf J=0)},$ which means that only the transversal part of ${\bf
  \Phi} $  may
enter the final free energy functional.

Derivation of the remaining terms in the free energy continues as
usual: the gradient expansion is continued up to terms second-order in 
${\bf \Phi}$, and first order in $\omega$. 
If we include, for the sake of generality, the possibility of 
a nonzero average component ($B_0$) to the magnetic field, we come to the free 
energy
\begin{eqnarray}
F[U] &=& \frac{\pi\nu}{8}\int d{\bf r}\, {\rm Str} (D[{\bf
  \Phi},\Lambda]^2
+ 2D_{xy} \tau_3\Lambda
\Phi_x^{\perp}\Phi_y^{\perp}
+2i\omega\Lambda Q ) 
\nonumber\\
&&
+ \frac{\pi}{2e}\int d{\bf r}J^{i}({\bf r}){\rm Str}(\Lambda \tau
_{3}\Phi ^{i}\left( {\bf r}\right) ),
\label{phifree}
\end{eqnarray}
where $D= v_F^2\tau/2$ and 
$D_{xy} = eB_0\tau D/mc$ (and we have assumed that the magnetic
field $B_0$ is classically weak, ie. $e B_0\tau/mc 
\ll 1$). 
The free energy may then by rewritten in terms of the $Q$-matrix, the
last term in Eq.~(\ref{phifree})
by means of a reexpression as a Wess-Zumino term\cite{Muz,Andreev}
(using the identity ${\rm div} {\bf J}=0$):
\begin{eqnarray}
F[Q] &=& \frac{\pi\nu}{8}\int d{\bf r} \,{\rm Str} (D(\nabla Q)^2 
 -D_{xy} \tau_3 Q[\nabla_x Q,\nabla_y Q]+2i\omega\Lambda Q)
\nonumber \\
&& + \frac{\pi}{8e}\int d{\bf r}\,{\bf J} \int_0^1 du \,{\rm Str}
\left(\tau_3 Q\left[\frac{\partial Q}{du},{\bf \nabla} Q \right]\right)
\label{phifr2}
\end{eqnarray}
The free energy (\ref{phifr2}) represents the most general free energy 
for electron diffusion in the presence of an external magnetic field
and nonzero local currents. 

The relation of the free energy (\ref{phifr2}) to the classical models of
diffusion in a field of random
velocities\cite{Derrida,KravLern,Fisher} is made clear if we consider
the correponding classical diffusion propagator. 
To derive this propagator, 
we consider $Q$ to be close to $\Lambda$ and expand
to quadratic order around $\Lambda$, via $Q= \Lambda (1+2i
P-2P^2+\ldots)$. The free energy reduces to
\begin{eqnarray}
F[P] = \frac{\pi\nu}{2}\int d{\bf r} \,{\rm Str}\left(P\left[
-D\nabla^2-i\omega - \frac{1}{e\nu}\Lambda \tau_3
{\bf J} \nabla\right]P\right).
\label{diffprop}
\end{eqnarray}
Comparison of the square brackets in Eq.~(\ref{diffprop}) with the
Fokker-Planck equation (\ref{dif1}), and use of the classical relation 
between the current and velocity,
\[
{\bf J}= e\nu {\bf v},
\]
shows that the free energy (\ref{phifr2})
coincides with the diffusion-advection model on the classical
level. The free energy (\ref{phifr2}) however contains additional
physics to the diffusion-advection model, since it describes also
quantum interference phenomena as represented by interaction of the
diffusion modes and contained in higher powers of $P$.
 
We now average over the random potentials. As we want in this section
only to make the connection with Refs. \cite{Derrida,KravLern,Fisher}, we do
not require a microscopic specification of the long-range disorder 
correlations, but may simply assume that the current
densities $J\left( {\bf r}\right) $ are random and delta-correlated,
so that
\begin{equation}
\langle J^{i}({\bf r})J^{j}({\bf r}^{\prime })\rangle =e^{2}G_{0}K^{ij}({\bf %
r}-{\bf r}^{\prime }),  \label{corrform}
\end{equation}
where $K^{ij}\left( {\bf r-r}^{\prime }\right) $ is defined as in 
Eq.~(\ref{b182}). In particular, $K^{ij}({\bf q})$, as given by 
Eq.~(\ref{dif2a}), remains nonzero in the limit of ${\bf q}\to 0$ and
takes the required transversal form.
The term
in the free energy, Eq.~(\ref{phifree}), that couples to the currents
becomes 
\begin{equation}
\delta F\rightarrow -\frac{\pi^2}{8e^{2}}\int d{\bf r}d{\bf r}^{\prime }\langle
J^{i}({\bf r})J^{j}({\bf r}^{\prime })\rangle {\rm Str}(\tau _{3}\Lambda
\Phi ^{i}\left( {\bf r}\right) ){\rm Str}(\tau _{3}\Lambda \Phi%
^{j}\left( {\bf r}^{\prime }\right) )  \label{dif6}
\end{equation}
In this way, we arrive at the free energy 
\begin{eqnarray}
F[Q] &=&\frac{\pi \nu }{8}\int \,{\rm Str}[D(\nabla Q({\bf r})
)^{2}
-D_{xy} \tau_3 Q[\nabla_x Q,\nabla_y Q]({\bf r}) 
+2i\omega \Lambda Q({\bf r})]d{\bf r}  \nonumber \\
&&-\frac{\beta }{16\pi }\int d{\bf r}d{\bf r}^{\prime }K^{ij}\left( {\bf r-r}%
^{\prime }\right) {\rm Str}\left( \tau _{3}\Lambda \Phi _{i}\left( {\bf r}%
\right) \right) {\rm Str}\left( \tau _{3}\Lambda \Phi _{j}\left( {\bf r}%
^{\prime }\right) \right),  \label{supfree}
\end{eqnarray}
where the dimensionless coefficient $\beta$ is defined by
\begin{equation}
\beta = 2\pi^3 G_{0}.  \label{dif7}
\end{equation}
An alternative reexpression of the free energy of Eq.~(\ref{supfree}) is
\begin{eqnarray}
F[Q] &=&\frac{\pi \nu }{8}\int {\rm Str}[D(\nabla Q\left( {\bf r}\right)
)^{2}
-D_{xy} \tau_3 Q[\nabla_x Q,\nabla_y Q]({\bf r}) 
+2i\omega \Lambda Q\left( {\bf r}\right)]d{\bf r}
 \nonumber \\
&&-\frac{2\beta }{\left( 32\pi \right) ^{2}}\int M\left( {\bf r}\right) \ln |%
{\bf r-r}^{\prime }|M\left( {\bf r}^{\prime }\right) d{\bf r}d{\bf r}%
^{\prime },  \label{supfr2}
\end{eqnarray}
where $M({\bf r})$ is defined as in Eq.~(\ref{Mdef}). 

So far we have specified the long-range disorder only through the
resultant current-current correlations, requiring that the correlator
(\ref{corrform})
remains nonzero in the limit of ${\bf q}\to 0$. Although such a
formulation has the advantage of generality, it is useful
to identify its relation to specific miscroscopic
models. It is clear from the
preceding sections that a delta-correlated RMF does not provide the
required form of current-current correlations. Instead, the magnetic
field correlations need to be of longer range: the Fourier transform
of the correlator $\langle B({\bf r})B({\bf r}^{\prime})\rangle$ must diverge
as $1/q^2$ as ${\bf q}\to 0$. 
 
We see that the
final term of Eq.~(\ref{supfr2}) is of the same form as that written
by Zhang and Arovas\cite{arovas}, although the latter work does not
contain a correct miscroscopic derivation; specifically, a proper
treatment along the lines of Ref.\cite{arovas}, since they deal with
short-range magnetic field correlations, would result in a
replacement of the function $K^{ij}({\bf q})$ appearing in the free energy of
Eq.~(\ref{supfree}) by $q^2 K^{ij}({\bf q})$. 
Indeed we may apply the
methods of this section to a model of short-range disorder plus a
delta-correlated RMF, to find just such a replacement
in the free energy of Eq.~(\ref{supfree}). It is
straightforward then to 
verify that this free energy reproduces
precisely the results of Wilke {\em et al}.\cite{Wilke} 
for the memory-effect correction to the conductivity in this model.

In contrast with the final
term of the free energy, Eq.~(\ref{b101}), for the delta-correlated
RMF problem, here the new term does contribute logarithmic
corrections to the conductivity. In the next
section we sum all such logarithmic corrections  by 
subjecting the free energy of Eq.~(\ref{supfr2}) to an
RG analysis.

%************************************************
%* Renormalisation Group                        *
%************************************************
\section{Renormalization Group Analysis}

\label{sec:RG} 
In this section we subject the free energy of Eqs.~(\ref{supfree})
and (\ref{supfr2}) to a  
renormalisation group analysis. In this way we 
find that the new term in the free energy (the
final term of Eq.~(\ref{supfr2})) leads to a new
contribution in the scaling form
of the diffusion coefficient with respect to frequency. The new
contribution is positive in sign, and hence competes with the usual negative 
contribution that represents weak localization. Moreover the new
contribution has precisely the same form as that which appears in
scaling form in the problem of classical 
superdiffusion\cite{Derrida,KravLern,Fisher}, according to Eq.~(\ref{dif3}).

Our approach is entirely analogous to the standard procedure applied in the
case of unitary disorder \cite{Brezin,Wegner,Efetov}. 
The only differences are caused by
the presence of the additional term in the free energy (\ref{supfree}). The
latter term leads to an extra `effective charge', $\beta $, whose flow is
coupled to that of the usual charge $t\equiv 8/(\pi \nu D)$.
In the standard case, the first order (one-loop) contribution in the
conventional unitary $\sigma $-model vanishes and the localizing behavior
originates from the two-loop diagrams that are the next order in $t^{-1}$.
For the free energy (\ref{supfree}), 
the $\beta $-term contributes to the effective charge $t$
already in the first order and therefore this contribution can be larger
than the conventional one. While the calculation of the conventional
two-loop diagrams is rather involved and needs a special regularization (a
dimensional regularization is usually used), the contribution of the $\beta $%
-term is much simpler and does not need any such 
regularization. 

Following a standard procedure (see e.g. Ref. \cite{Efetov}) we separate the
unitary supermatrix $U({\bf r})$ into slow $\tilde{U}({\bf r})$ and fast $%
U_{0}({\bf r})$ parts: 
\begin{equation}
U({\bf r})=\tilde{U}({\bf r})U_{0}({\bf r}).  \label{Usep}
\end{equation}

The fast part $U_{0}\left( {\bf r}\right) $ contains in the momentum
representation momenta in the interval $\lambda <k<k_{0}$, where $k_{0}$ is
the upper cutoff momentum. To guarantee that the rotational symmetry is not
violated after this integration we impose an infrared cutoff in the momentum
integrals by adding the following term in the free energy 
\[
F_{reg}=-\frac{2\lambda ^{2}}{t}\int {\rm Str}\left( \Lambda Q\right) d{\bf %
r,} 
\]
where $\lambda ^{2}\gg \tilde{\omega}\equiv \omega /D$.

Substituting the parametrization, Eq. (\ref{Usep}), into the free energy (%
\ref{supfree}), we find $F=F^{t}+F^{\beta }$, 
\begin{eqnarray}
F^{t} &=&\frac{1}{t}\int d{\bf r}{\rm Str}\left[ (\nabla
Q_{0})^{2}+2[Q_{0},\nabla Q_{0}]\tilde{\Phi}+[Q_{0},\tilde{\Phi}]^{2}\right.
\label{c2} \\
&&\left. \hspace{3cm}+2i\tilde{\omega}\overline{\tilde{U}}\Lambda \tilde{U}%
Q_{0}\right] ,  \nonumber \\
F^{\beta } &=&-\frac{\beta }{16\pi }\int d{\bf r}d{\bf r}^{\prime
}K^{ij}\left( {\bf r-r}^{\prime }\right) \left[ {\rm Str}(\tau
_{3}Q_{0}\left( {\bf r}\right) \tilde{\Phi}_{i}\left( {\bf r}\right) ){\rm %
Str}(\tau _{3}Q_{0}\left( {\bf r}^{\prime }\right) \tilde{\Phi}_{j}\left( 
{\bf r}^{\prime }\right) )\right.  \nonumber \\
&&\left. \hspace{2cm}
+{\rm Str}(\tau _{3}\Lambda \Phi _{0i}\left( {\bf r}%
\right) ){\rm Str}(\tau _{3}\Lambda \Phi _{0j}\left( {\bf r}^{\prime
}\right) )
+2\,{\rm Str}(\tau _{3}Q_{0}\left( {\bf r}\right) \tilde{\Phi}%
_{i}\left( {\bf r}\right) ){\rm Str}(\tau _{3}\Lambda \Phi _{0j}\left( {\bf r%
}^{\prime }\right))\right] ,  \nonumber
\end{eqnarray}
where ${\bf \tilde{\Phi}}=\overline{\tilde{U}}{\bf \nabla }\tilde{U}$, 
${\bf \Phi }%
_{0}=\bar{U}_{0}{\bf \nabla }U_{0}$ and $Q_{0}=U_{0}\Lambda \bar{U}_{0}$.
Notice we do not include any contribution from 
the topological term (that couples to $D_{xy}$) in the free energy of 
Eq.~(\ref{supfree}), since as usual this term
does not contribute to the RG flow equations to any perturbative order.

Our task is now to integrate $\exp \left( -F\right) $ over $U_{0}$ using
perturbation theory and obtain a new free energy functional $\widetilde{F}$
containing second powers of $\widetilde{\Phi }_{\alpha }\left( {\bf r}%
\right) $. We parametrize $U_{0}$ by 
\begin{equation}
U_{0}=\left( \frac{1-iP}{1+iP}\right) ^{1/2},  \label{c3}
\end{equation}
expand the free energy to the required order in $P$ and average over $P$ to
obtain the required renormalized free energy.

It is important to mention that the bare free energy, which is 
quadratic in $P$ and with which the subsequent averaging will be done, 
originates from $F^{t}$. As
concerns the contribution from $F^{\beta }$, its expansion in $P$ begins
with $P^{4}$ terms. Therefore the new term does not modify the diffusion
equation itself but changes interaction between the diffusion modes.

Terms coming to the renormalized functional from $F^{t}$ are well known and
we simply use the results written in Ref. \cite{Efetov}. It is important
that one does not obtain from $F^{t}$ any contribution to the renormalized $%
\beta $-term. \ As concerns $F^{\beta }$\bigskip , we write $F^{\beta
}=F_{1}^{\beta }+\tilde{F}^{\beta }$, where 
\begin{eqnarray}
F_{1}^{\beta } &=&\frac{\beta }{4\pi }\int d{\bf r}d{\bf r}^{\prime
}K^{ij}\left( {\bf r-r}^{\prime }\right) \{{\rm Str}(\tau _{3}\Lambda
P\left( {\bf r}\right) \tilde{\Phi}_{i}\left( {\bf r}\right) ){\rm Str}(\tau
_{3}\Lambda P\left( {\bf r}^{\prime }\right) \tilde{\Phi}_{j}\left( {\bf r}%
^{\prime }\right) )  \label{c3a} \\
&&+{\rm Str}(\tau _{3}\Lambda \tilde{\Phi}_{i}\left( {\bf r}\right) ){\rm Str%
}(\tau _{3}\Lambda P^{2}\left( {\bf r}^{\prime }\right) \tilde{\Phi}%
_{j}\left( {\bf r}^{\prime }\right) )+{\rm Str}(\tau _{3}\Lambda \tilde{\Phi}%
_{i}\left( {\bf r}\right) ){\rm Str}(\tau _{3}\Lambda \nabla _{r^{\prime
}j}P\left( {\bf r}^{\prime }\right) P\left( {\bf r}^{\prime }\right) )\}, 
\nonumber \\
\tilde{F}^{\beta } &=&-\frac{\beta }{16\pi }\int d{\bf r}d{\bf r}^{\prime}
K^{ij}({\bf r}-{\bf r}^{\prime})
{\rm Str}(\tau
_{3}\Lambda \tilde{\Phi}_i({\bf r})) 
{\rm Str}(\tau
_{3}\Lambda \tilde{\Phi}_j({\bf r}^{\prime}))
.  \nonumber
\end{eqnarray}
Then we integrate over $P$ using the contraction rule \cite{Efetov} 
\begin{equation}
<{\rm Str}\left( PM_{1}\right) {\rm Str}\left( PM_{2}\right) >_{0}=\frac{t}{%
8\left( k^{2}+\lambda ^{2}\right) }{\rm Str}\left( M_{1}M_{2}\right) \text{,}
\label{c4}
\end{equation}
and $<PP>=0$, where the supermatrices $M_{1}$ and $M_{2}$ anticommute with $%
\Lambda $ and are self-conjugate.

We see that in the first order approximation, there is no contribution to
the renormalized $\beta $-term coming from $F^{\beta }$. So, we conclude
that in the one-loop order the $\beta $-term is not renormalized and the
coefficient $\beta $ keeps its bare value.

The first order contribution to the diffusion (first) term in the action,
Eq. (\ref{b100}), comes from only the first term in $F_{1}^{\beta }$ in Eq. (%
\ref{c3a}), while the contributions from the second and the third terms
in Eq.~(\ref{c3a})
vanish. The contribution from
the first term in $F_{1}^{\beta }$ contains the integral 
\begin{equation}
-\frac{\beta t}{32\pi }\int {\rm Str}\left( \Phi _{i}^{\perp }\left( {\bf r}%
\right) \Phi _{j}^{\perp }\left( {\bf r}\right) \right) d{\bf r}\int 
\frac{d^2{\bf k}}{(2\pi)^2} \left(
\delta ^{ij}-\frac{k^{i}k^{j}}{k^{2}}\right)\frac{1} 
{k^{2}+\lambda^{2}}.  \label{c5}
\end{equation}
After performing the above integral, 
the renormalized free energy $\tilde{F}$ can be written in $%
d=2+\varepsilon $ dimensions (for small $\varepsilon $) as
\begin{eqnarray}
\tilde{F} &=&\frac{1}{t}\int d{\bf r}\left[ {\rm Str}(\nabla \tilde{Q}%
)^{2}\left\{ 1+\frac{t^{2}}{64}\left( \frac{1}{2}-\frac{1}{d}\right)
I_{d}^{2}+\frac{\beta t^{2}}{256\pi }I_{2}\right\} +2i\tilde{\omega}{\rm Str}%
(\Lambda \tilde{Q})\right]  \nonumber \\
&&\hspace{4cm}
-\frac{\beta }{16\pi }\int d{\bf r}d{\bf r}^{\prime}
K^{ij}({\bf r}-{\bf r}^{\prime})
{\rm Str}(\tau
_{3}\Lambda \tilde{\Phi}_i({\bf r})) 
{\rm Str}(\tau
_{3}\Lambda \tilde{\Phi}_j({\bf r}^{\prime})).
\label{trenorm}
\end{eqnarray}

The second term in the figure brackets in Eq. (\ref{trenorm}) is the
conventional two-loop contribution of the unitary $\sigma $-model. To write
it properly one may use a dimensional regularization, which
is the reason this term is written 
for $\varepsilon \neq 0$. At the same time, there are no
ambiguities in the calculation of the third term in the figure brackets and it
is written for $\varepsilon \ll 1$ already in $2D$.

The Gell-Man-Low function for 
\[
\tilde{t}=\frac{t}{16\pi } 
\]
becomes, after continuation to $2$ dimensions, 
\begin{eqnarray}
\frac{d\tilde{t}}{d\ln \lambda } &=&\frac{1}{2}\tilde{t}^{3}(\beta -1)+O(%
\tilde{t}^{4}).  \label{c6} \\
\beta &=&const  \nonumber
\end{eqnarray}

The solution to the flow equation, Eq. (\ref{c6}) may be found by
integrating over $\lambda $ up to the ultraviolet cutoff of $1/l_{{\rm tr}}$,
and we obtain 
\begin{equation}
\tilde{t}(\omega )=\tilde{t}_{0}
\left[1+\frac{1}{2}(\beta -1)\tilde{t}_{0}^{2}\ln
\left(\frac{1}{\omega \tau_{{\rm tr}} }\right)\right]^{-1/2},  \label{c7}
\end{equation}
where $\tilde{t}_{0}$ is the bare value of $t$ proportional to the classical
resistivity. Rewriting Eq.~(\ref{c7}) in terms of the diffusion
coefficient and reintroducing $G_0$ (see Eq.~(\ref{corrform})), we find
\begin{eqnarray}
D(\omega) = D_0\left[1+\left\{\frac{G_0}{4\pi\nu^2
        D_0^2}-\frac{1}{8\pi^4\nu^2 D_0^2}\right\}\ln
\left(\frac{1}{\omega \tau_{{\rm tr}} }\right)\right]^{1/2}.
\label{Dscale}
\end{eqnarray}
The classical relation ${\bf J}= e\nu {\bf v}$ between
the current density ${\bf J}$ and the 
velocity of a single particle, ${\bf v}$, leads to 
$G_{0}=\nu ^{2}\gamma _{0}$.
The coefficient of the positive correction to the diffusion
coefficient can then be reexpressed as
\begin{eqnarray*}
\frac{G_0}{4\pi\nu^2 D_0^2} = g_0,
\end{eqnarray*}
where $g_0 = \gamma_0/4\pi D_0^2$. Comparison with the scaling equation
(\ref{dif3}) shows that this coefficient coincides precisely with
that found previously\cite{KravLern,Fisher}
in the scaling form for the problem of classical superdiffusion. 

\begin{figure}[tbp]
\begin{center}
\epsfig{file=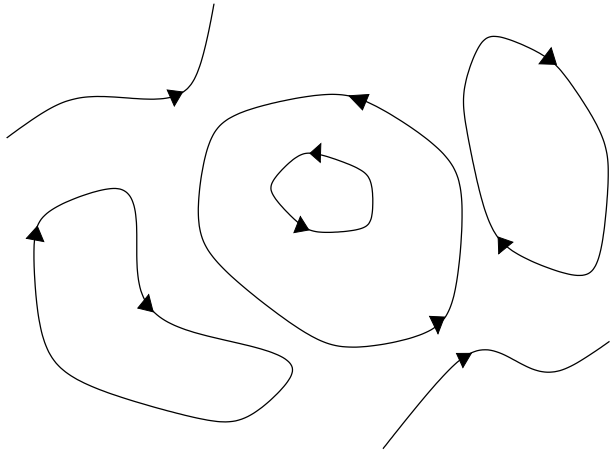,height=6cm} 
\caption{\label{fig:cycles}
Cycles of local current along which semiballistic propagation is
possible.}
\end{center}
\end{figure}

This correspondence provides immediately 
an classical physical interpretation for the positive part of the 
correction to the diffusion
coefficient in Eq.~(\ref{Dscale}).
For the flow to be
correlated according to Eqs.~(\ref{dif2}) and (\ref{corrform}),
random stationary flow `cycles' should exist
in the system.   A tendency towards ballistic motion along the cycles of
local current (see Fig.\ref{fig:cycles}) leads to the positive, or
superdiffusive,
correction to the diffusion coefficient.

%***************************************
%* Figure: cycles                      *
%***************************************

The difference of the electron model we consider here
with the models of classical
superdiffusion\cite{Derrida,KravLern,Fisher} 
is that for the electron model quantum
interference processes are also important and give rise to 
negative corrections to the diffusion coefficient. This is the negative
contribution in the scaling form of 
Eq.(\ref{Dscale}), and it coincides with the
usual contribution known for the unitary
ensemble for short-range disorder\cite{Brezin,Wegner}. 

The presence of both classical and quantum corrections to the
conductivity, which are both logarithmic and therefore in competition
with each other, provides the interesting
possibility of a quantum phase transition in two dimensions at weak
disorder (at $\beta=1$), 
tuned by the relative strengths of the short-range and
long-range disorder. 

In the case when the positive contribution in the scaling form
 for the diffusion coefficient
dominates over the negative contribution ($\beta>1$), 
Eq. (\ref{c7}) can be used for an arbitrarily low
frequencies as soon as $\tilde{t}_{0}\ll 1$. This is 
because the effective charge
becomes even smaller when $\omega \rightarrow 0$ and one may safely neglect
higher order terms in Eq. (\ref{c6}\.{)}. This type of behavior is similar
to that known in the symplectic symmetry class that describes short-range,
spin-orbit impurities\cite{hikami}, although the final forms of the scaling
functions in the two systems are different.
In such cases, the type of the renormalization group equations is known as a
`zero-charge' situation. The possibility to use Eq. (\ref{c7}) down to zero
frequencies means that this represents the complete solution. This
contrasts with, for example, 
the localization behavior for the orthogonal ensemble where
the effective charge (resistivity) $\tilde{t}$ $\ $grows when decreasing the
frequency and the solution of the RG equations loses its applicability as $%
\tilde{t}$ becomes of the order of unity. 

The result for the scaling of the diffusion coefficient with frequency,
Eqs.~(\ref{Dscale}), 
may be obtained also perturbatively order-by-order,
by calculating the conductivity
directly, as in section \ref{sec:direct}. In such a calculation a
correction, $\delta\sigma^{ij}$, appears that is 
similar to $R_c^{ij}$ of Eq.(\ref{Rceq}): it is
proportional to the current-current correlator and has no
counterpart in the conventional calculation\cite{Lerner,Efetov}
 for short-range disorder:
\begin{eqnarray}
\delta\sigma^{ij} = \frac{\pi}{16}\int d({\bf r}-{\bf r}^{\prime})
{\rm Str}(k(1-\tau_3)Q^{12}({\bf r}^{\prime})k(1-\tau_3) Q^{21}({\bf
  r}))\langle J^i({\bf r})J^j({\bf r}')\rangle.
\label{condir}
\end{eqnarray}
Expanding each $Q$ in the 
right-hand side of Eq.~(\ref{condir}) to first-order in $P$, where $Q=
\Lambda(1+iP)(1-iP)^{-1}$, and averaging over $P$ leads to a
correction to the conductivity that corresponds to the positive
contribution in the scaling form for
 the diffusion coefficient of
Eq.~(\ref{Dscale}), after a perturbative expansion. The 
correction to the conductivity corresponding to the negative contribution in 
Eq.~(\ref{Dscale}) can also be derived diagrammatically in the usual
way\cite{Lerner,Efetov}. 
The equivalence of the direct calculation
 of the conductivity with the RG approach for the diffusion coefficient
follows naturally from the Einstein
relation, which must hold for the electron problem. 
There does not seem to be a connection between
the mobility studied in Refs.~\cite{Derrida,KravLern,Fisher} 
and the conductivity of the
electron gas with long-range disorder, and its
dependence on the frequency differs from that for the diffusion coefficient.

In summary, in this section we have subjected the free energy
(\ref{supfr2}) to a RG procedure to derive to scaling equation for the 
diffusion coefficient. As showed in section \ref{sec:superdiff},
the free energy (\ref{supfr2}) represents a quantum
generalisation of the diffusion-advection problem and applies when the 
current-current correlator due to long-range disorder remains finite
in the limit of ${\bf q}\to 0$. The scaling form
for the diffusion
coefficient contains a positive contribution, which corresponds to classical
superdiffusion, and a negative contribution, which is the usual weak
localization correction for the unitary ensemble. 
Since they are of opposite sign, the two contributions
may compete and this leads to the
possibility of a quantum phase transition in two dimensions at weak
disorder, tuned by the relative disorder strengths. 

\section{Summary and Discussion}

\label{sec:discuss} 
In this paper, we have studied the 2D electron gas 
with long range disorder and broken time-reversal symmetry. 
We have provided a careful derivation of the free energy for this
system. We have shown how the 
the ballistic $\sigma$-model may be derived in a controlled manner for 
this problem, due to the presence of the small parameter
$\tau/\tau_{\rm tr}$, thus providing the first formal derivation of
this model. We also derived certain additional terms which are beyond the
usual form of the ballistic $\sigma$-model but  become important for
shorter-range disorder. Integration over nonzero harmonics of the
$Q$-matrix leads to a free energy with
with an extra term beyond those of the conventional unitary
$\sigma$-model. The  evaluation of the extra term requires the use of a proper
regularisation procedure of the ballistic $\sigma$-model
that ensures the preservation of gauge invariance and that has been
verified by diagrammatic means. 

This demonstrates that, within the
scope of our analysis of the delta-correlated RMF model, 
the extra term in the free energy 
seems to remain irrelevant in the
calculation of corrections to the conductivity that are logarithmic in
frequency. As a result we do not find
evidence for deviation of the localization behaviour
of the delta-correlated random magnetic field model from that of the
conventional unitary ensemble.  As such our conclusions coincide
with those of the original work of
Aronov {\em et al}., although for
considerably more subtle reasons than were originally realised in that 
work (which remains a partial analysis due to their neglect of the
higher harmonics of the $Q$-matrix). 

Certain questions for this model remain 
regarding the role of the massive (a) modes, which 
describe fluctuations of the eigenvalues of $Q$ and the Cooperons,
and which have been neglected in our analysis. For example, the
precise value of the single-particle lifetime that couples to the
massive modes (a) remains to be determined, and is likely to remain finite 
as the infrared cutoff $\kappa$ is taken to zero. This suggests that
an improved saddle-point may be found which does not contain any
divergencies in the limit of $\kappa\to 0$. A proper treatment
of the massive (a) modes would allow the application of the ballistic
$\sigma$-model down to distances of the order of the single-particle mean free
path, which is desirable since the evaluation of the new term in the
free energy depends crucially on the 
the regularization procedure at such short length-scales.

The derivation of the free energy was supplemented by direct perturbative
calculations within the field theory formalism. We calculated the
conductivity directly and showed the appearance of an additional
correction, that is  
analogous to the new term in the free energy and is higher order than 
logarithmic. We also
calculated the current-current correlation function, thus providing a
physical interpretation for the coefficient of the new term in the
free energy.  

We then generalised the discussion to include the possibility of
longer ranged disorder. We derived the free energy for a model 
containing both
short-range disorder and longer-ranged disorder, and showed how the
long-range disorder couples to the free energy via a Wess-Zumino
term. If the long-range disorder is sufficiently
long-ranged that the correlator of local currents remains finite in the
limit of ${\bf q}\to 0$, then a new term appears in the free energy
which is responsible for new corrections to the
conductivity that are logarithmic in frequency. 
We have demonstrated how this free energy may be
described as a quantum generalisation of the
diffusion-advection model. By performing a RG analysis of the free
energy, we have shown how two competing contributions to the scaling 
form for the diffusion
coefficient appear. The positive contribution corresponds to classical
superdiffusion in the presence of long-ranged currents, while the
negative contribution is the conventional, weak localization correction
for the unitary ensemble. The competition between the classical and
quantum contributions to the conductivity 
leads to the interesting possibility of a quantum phase 
transition in two dimensions at weak
disorder, tuned by the relative disorder strengths. 

We remark that a direct 
diagrammatic calculation of the leading logarithmic correction to
the conductivity would be a useful check of the predictions of the field 
theory for the delta-correlated RMF model. The diagrammatic arguments
of  Gornyi 
{\em et al}.\cite{Gornyi} do not fulfill the purpose of such a check:
although their calculation of the current-current correlation function 
seems valid, their calculation of the conductivity relies on a
relation between the conductivity and the current-current correlation
function that seems questionable for the delta-correlated RMF model,
since it depends on the quantity $\tau$ that
formally vanishes within the SCBA. Instead their arguments
would seem to be applicable when the RMF is a weak 
perturbation to sufficiently strong, short-range
disorder, in which case a well-defined $\tau$ does exist. A
direct diagrammatic calculation of the first logarithmic correction to the
conductivity, which is not simple even for short-range
disorder, therefore remains to be attempted for the delta-correlated 
RMF model. 

We remark also that further 
analysis of the free energy, (\ref{supfr2}), with its new
term, is deserved. For example, the influence of the new
term on the scaling of the conductivity at small conductance, 
and its interplay with the
usual topological term, remains an interesting open question. The 
general expression (\ref{phifr2}) for the free energy of
a disordered conductor in the presence of equilibrium currents also
presents
new possibilities for further investigation, 
possibly in different dimensionalities. 

The method of derivation of the ballistic $\sigma$-model in this
paper, by which a controlled expansion of the free energy becomes
possible through the smallness of the parameter $\tau/\tau_{\rm tr}$,
opens a promising avenue for further development of ballistic 
field-theories. It is possible that progress in this direction
may help to resolve the problem of ``repetitions''\cite{BogKeat} 
in the ballistic 
$\sigma$-model for chaotic systems.

We are grateful to I. Aleiner, A. Altland, B.L. Altshuler, V. E. Kravtsov,
A. I. Larkin, M. V. Feigel'man, B. D. Simons, 
A. Tsvelik, and M. Zirnbauer for helpful
discussions during the course of this work and gratefully acknowledge the
financial support of the {\it `Sonderforschungsbereich'} 237, 491 and of the 
{\it Schwerpunktprogramm ``Quanten Hall Systeme''.}

%
%\end 

%******************************************************
%* APPENDIX                                           *
%******************************************************
\appendix

%*******************************************************
%* A-squared term                                      *
%*******************************************************

\section{Neglect of $A^2$ term}

\label{app:asquare} In this appendix we justify more carefully the neglect
of the $A^2$ term in the original Lagrangian, Eq.~(\ref{Lag0}). In fact this
approximation is standard and has been used in previous analytic works on
the subject\cite{Aronov,miller}. Indeed our approach applies also
to the case of 
long-range impurities (scalar disorder) for which $A^2$-like terms are
completely absent.

Even so we investigate here more carefully the influence of the $A^{2}$
term. In fact its neglect was made only for convenience and is not necessary
in the early stages of the calculation: instead, it may be included exactly
in the disorder average. In doing so, ${\cal L}_{{\rm int}}$ of Eq.~(\ref
{Lint}) is replaced by 
\[
{\cal L}_{{\rm int}}=\int \bar{\psi}^{\alpha }({\bf r})\psi ^{\beta }({\bf r}%
^{\prime })\tau _{3\,\beta \beta }k_{\beta \beta }\widehat{J}_{{\bf r},{\bf r%
}^{\prime }}\bar{\psi}^{\beta }({\bf r}^{\prime })\psi ^{\alpha }({\bf r}%
)\tau _{3\,\alpha \alpha },
\]
where 
\[
\widehat{J}_{{\bf r},{\bf r}^{\prime }}\equiv -\frac{1}{2}\sum_{ij}\nabla _{%
{\bf r}-{\bf r}^{\prime }}^{i}[V^{-1}({\bf r}-{\bf r}^{\prime })-4im\delta (%
{\bf r}-{\bf r}^{\prime })\openone\bar{\psi}({\bf r})\psi ({\bf r}%
)]_{ij}^{-1}\nabla _{{\bf r}-{\bf r}^{\prime }}^{j},
\]
where $\openone$ is a unit matrix in $ij$-space. The decoupling of the
interaction term, ${\cal L}_{{\rm int}}$, by a $Q$-matrix follows as before,
as does the integration over the $\psi $-fields. The Lagrangian of Eq.~(\ref
{Lag1}) is now replaced by 
\begin{eqnarray}
{\cal L} &=&\int \left[ -\frac{1}{2}{\rm Str}\ln \left[ i{\cal H}_{0}({\bf r}%
)\delta ({\bf r}-{\bf r}^{\prime })+\widehat{V}_{{\bf r},{\bf r}^{\prime }}%
\widetilde{Q}({\bf r},{\bf r}^{\prime })\right. \right.   \nonumber \\
&&\left. \left. \mbox{}\qquad +\int d{\bf r}^{\prime \prime }d{\bf r}%
^{\prime \prime \prime }{\rm Str}\left( Q({\bf r}^{\prime \prime \prime },%
{\bf r}^{\prime })V_{{\bf r}^{\prime \prime \prime },{\bf r}^{\prime \prime
}}^{ik}V_{{\bf r}^{\prime \prime },{\bf r}^{\prime }}^{kj}\nabla _{{\bf %
r^{\prime \prime \prime }}-{\bf r}^{\prime }}^{i}\nabla _{{\bf r^{\prime
\prime \prime }}-{\bf r}^{\prime }}^{j}Q({\bf r}^{\prime },{\bf r}^{\prime
\prime \prime })\right) \right] \right.   \nonumber \\
&&\left. \mbox{}\qquad +\frac{1}{4}{\rm Str}\left( Q({\bf r},{\bf r}^{\prime
})\widehat{V}_{{\bf r},{\bf r}^{\prime }}Q({\bf r}^{\prime },{\bf r})\right) %
\right] d{\bf r}d{\bf r}^{\prime }.  \label{Lagex}
\end{eqnarray}
We see that an additional term has arisen inside the logarithm (second line
of Eq.~(\ref{Lagex})). This term is second-order in the scattering
amplitude, $V^{ij}$, and acts to renormalize the chemical potential. Indeed,
due to the presence of the ${\rm Str}$ operating on this term, we see this
term has no effect on the saddle-point $Q_{p}=\Lambda _{p}$, since the
latter is supersymmetric.

Consider now fluctuations around the saddle-point, of the form parametrized
by Eq.~(\ref{param}) (such that $Q_{{\bf p}}^{2}=1$). The unusual feature of
the additional term inside the logarithm of Eq.~(\ref{Lagex}) is that, due
to the presence of the ${\rm Str}$ operator, it commutes with the rotation
matrices $U$ and $V_{{\bf n}}$. Performing an expansion of the logarithm in
this term will therefore not give any additional contribution to the free
energy to any perturbative order. Instead, the effect of the additional term
is comparable to that of the (a)-modes, corresponding to fluctuations in $Q_{%
{\bf p}}^{2}$, and hence may be neglected consistently.

An alternative argument for the neglect of the $A^2$ term may be formulated
as follows: we may treat the $A^2$ term in the Lagrangian (\ref{Lag0})
perturbatively and check that it gives only a parametrically small
correction to the previously-derived terms in the free energy.

A leading-order contribution of the $A^{2}$ term, that is fourth-order in $A$%
, may be estimated via a third-order cumulant expansion: this gives rise to
the contribution to the Lagrangian of 
\begin{eqnarray*}
\delta {\cal L}_{1} &\sim &\int d{\bf r}d{\bf r^{\prime }}d{\bf r}^{\prime
\prime }\left\langle \left( \bar{\psi}({\bf r})\frac{e}{mc}{\bf A}({\bf r}).{%
\nabla }_{{\bf r}}\tau _{3}\psi ({\bf r})\right) \left( \bar{\psi}({\bf r}%
^{\prime })\frac{e}{mc}{\bf A}({\bf r}^{\prime }){\nabla }_{{\bf r}^{\prime
}}\tau _{3}\psi ({\bf r}^{\prime })\right) \times  \right.\\
&&\left.\hspace{2cm}\times \left( \bar{\psi}({\bf r}^{\prime \prime })\frac{e^{2}}{%
mc^{2}}A^{2}({\bf r}^{\prime \prime })\psi ({\bf r}^{\prime \prime })\right)
\right\rangle 
\end{eqnarray*}
Choosing a typical pairing of the $A$ and $\psi $ fields gives an estimate
of a typical resulting contribution to the free energy: 
\[
\delta F_{1}\sim \frac{1}{\epsilon _{F}^{2}}\int d{\bf r}\sum_{{\bf n}_{1},%
{\bf n}_{2},{\bf n}_{3}}V_{p_{F}({\bf n}_{2}-{\bf n}_{1})}^{ik}V_{p_{F}({\bf %
n}_{3}-{\bf n}_{1})}^{kj}n_{1}^{i}n_{1}^{j}{\rm Str}(Q_{{\bf n}_{1}}({\bf r}%
)Q_{{\bf n}_{2}}({\bf r})Q_{{\bf n}_{3}}({\bf r}))
\]
In fact due to the combination of three $Q$'s in the structure of $\delta
F_{1}$, this terms does not contribute at all to the free energy of the form
of Eq.~(\ref{Lag3}), upon expanding in $P_{m}$. In addition we note that it
is second-order in the scattering amplitude $V^{ij}$: for this reason it may
be checked that in any case it appears with a small factor $(\epsilon
_{F}\tau _{{\rm tr}})^{-1}\ll 1$ with respect to the usual collision term
(see e.g. Eq.~(\ref{M+K})).

The other leading-order contribution, that is fourth-order in $A$, appears
from the second-order cumulant: 
\[
\delta {\cal L}_{2}\sim \int d{\bf r}d{\bf r^{\prime }}\left\langle 
\left( \bar{%
\psi}({\bf r})\frac{e^{2}}{mc^{2}}A^{2}({\bf r})\psi ({\bf r})\right) \left( 
\bar{\psi}({\bf r}^{\prime })\frac{e^{2}}{mc^{2}}A^{2}({\bf r}^{\prime
})\psi ({\bf r}^{\prime })\right)\right\rangle .
\]
Choosing a typical pairing as before gives an estimate of a typical
resulting contribution to the free energy as 
\begin{eqnarray*}
\delta F_{2} &\sim &\frac{1}{\epsilon _{F}^{2}}\int d{\bf r}\sum_{{\bf n}%
_{1},{\bf n}_{2}}\tilde{w}_{{\bf n}_{1}-{\bf n}_{2}}{\rm Str}(Q_{{\bf n}%
_{1}}({\bf r})Q_{{\bf n}_{2}}({\bf r})), \\
\tilde{w}_{{\bf n}_{1}-{\bf n}_{2}} &=&p_{F}^{2}\int d^2{\bf p}_{3}V_{{\bf p}%
_{3}}^{ik}\,V_{{\bf p_{3}}+p_{F}({\bf n}_{1}-{\bf n}_{2})}^{ki}.
\end{eqnarray*}
We notice that the term $\delta F_{2}$ is again second-order in the
scattering amplitude $V^{ij}$, as compared to the usual collision term in
the ballistic $\sigma $-model (Eq.~(\ref{M+K})) which is only first-order in 
$V^{ij}$. As a result it is straightforward to check that the term $\delta
F_{2}$ produces a renormalization of the collision term by a parametrically
small factor $(\epsilon _{F}\tau _{{\rm tr}})^{-1}\ll 1$. Higher-order
cumulants of the $A^{2}$ term in the original Lagrangian will clearly
produce only smaller renormalizations of the collision term corresponding to
higher powers of this small parameter. In this way we conclude that it is
justified to neglect the $A^{2}$ term as claimed in the main text.

\end{document}